\pdfoutput=1
\documentclass[10pt]{iopart}
\usepackage{graphicx}
\usepackage{url}

\def\sun{\odot}
\def\h0units{\mathrm{km\,s^{-1}\,Mpc^{-1}}}
\def\cunits{\mathrm{km\,s^{-1}}}
\newcommand{\om}{\Omega_{\rm M}}
\newcommand{\ok}{\Omega_K}
\newcommand{\ola}{\Omega_{\Lambda}}

\newcommand{\ml}{d_{L,m,2,2}}
\def\aap{A\&A\,  }

\def\apj{ApJ\,  }
\def\apjl{ApJ\,  }
\def\baas{Bulletin of the American Astronomical Society}

\def\mnras{MNRAS\,  }

\def\nat{Nature\,  }
\def\pasj{PASJ\,  }
\def\pasp{PASP  }
\def\pasa{PASA  }

\begin{document}
\title
{
The giants arcs as modeled by the superbubbles
}
\vskip  1cm
\author     {Lorenzo  Zaninetti}
\address    {Physics Department,
 via P.Giuria 1,\\ I-10125 Turin,Italy }
\ead {zaninetti@ph.unito.it}
\begin {abstract}
The giant arcs in the clusters of galaxies
are modeled in the framework of the superbubbles.
The density of the intracluster medium is assumed 
to follow a  hyperbolic behavior.
The analytical law of motion is function of the elapsed time
and the polar angle.
As a consequence the flux of kinetic energy
in the expanding thin layer decreases
with increasing polar angle
making the giant arc invisible to the astronomical observations.
In order to  calibrate the arcsec-parsec   conversion
three cosmologies are analyzed.
\end{abstract}
\vspace{2pc}

\maketitle

\section{Introduction}

The giant  arcs in the cluster of galaxies 
start to be observed as narrow-like shape by
\cite{Lynds1986,Paczynski1987,Soucail1987} and
the first theoretical explanation 
was the gravitational lensing, see
\cite{Kovner1987,Waldrop1987,Soucail1988,Narasimha1988}.
The determination of the statistical parameters
of the giant arcs has been analyzed 
in order to derive the cosmological parameters, see \cite{Kaufmann2000},
in connection with the $\Lambda$CDM cosmology, see \cite{Wambsganss2004},
in the framework of the triaxiality and
substructure of CDM halo, see \cite{Dalal2004},
in connection with the 
Wilkinson Microwave Anisotropy Probe (WMAP) data, see \cite{Li2006},
including the effects of baryon cooling  in dark matter N-body  
simulations, see \cite{Wambsganss2008} and 
in order to derive the photometric properties of 105 giant arcs that
in the Second Red-Sequence Cluster Survey (RCS-2), see \cite{Bayliss2012}.
The gravitational lensing is the most common  theoretical 
explanation, we select some approaches among others:
\cite{Hammer1989} evaluated the mass
distribution inside distant clusters,
\cite{Wu1996} studied  the statistics of giant arcs
in flat cosmologies with and without a cosmological constant,
\cite{Lewis2001} analyzed how the gravitational lensing
influences  the surface brightness of giant luminous arcs
and 
\cite{Mahdi2014} 
used  the warm dark matter (WDM) cosmologies  to explain  the lensing
in  galaxy clusters.
Another theoretical line of research  explains the
giant arcs as shells originated  by the galaxies in the
cluster: \cite{Dekel1988,Braun1988} analyzed the 
the limb-brightened shell model, the gravitational lens model and the
echo model, 
\cite{Efremov1998} suggested that the Gamma-ray burst (GRB) 
explosions are the sources of the   shells with sizes
of many kpc,
and \cite{Zaninetti2017c}  suggested a connection 
between the Einstein ring associated  to SDP.81
and  the evolution of a superbubble (SB)
in the intracluster medium.

This paper analyzes in Section~\ref{sec_cosmology} 
three cosmologies in
order to calibrate 
the transversal distance  which  allows to convert
the arcsec in pc.
Section~\ref{sec_sb}  is devoted  to the evolution
of a SB  in the intracluster medium.
Section~\ref{sec_astro} reports the observations
of the giant arcs and the first phase of a SB.
Section~\ref{sec_simulation} reports 
the various steps which allow  to reproduce 
the shape of the giant arc A2267
and the multiple arcs visible in the
cluster of galaxies.
Section~\ref{imagetheory} is dedicated to 
theory of the image:  
analytical formulae 
 explain the hole in the central part of the SBs
and numerical results reproduce the details
of a giant arc.

\section{Adopted cosmologies}
\label{sec_cosmology}

In the following we review three cosmological
theories.
\subsection{$\Lambda$CDM cosmology}

The  basic parameters of $\Lambda$CDM cosmology
are: 
the Hubble constant, $H_0$, expressed in  $\h0units$,
the velocity of light, $c$,  expressed in $\cunits$, and
the three numbers $\om$, $\ok$, and $\ola$,
see \cite{Zaninetti2016a} for more details.
In the case of  the Union 2.1 compilation, see
\cite{Suzuki2012},
the parameters are  $H_0 = 69.81 \h0units$, $\om=0.239$  and  $\ola=0.651$.
To have the  luminosity distance, $D_L(z;H_0,c,\om,\ola)$, 
as a  function of the redshift only, 
we apply     the minimax rational approximation,
which is characterized by  two parameters, $p$ and $q$.
The luminosity distance, $D_{L,3,2}$,
when $p=3$ and $q=2$ 
\begin{eqnarray}
D_{L,3,2}= 
\frac
{
- 7.761- 1788.53\,z- 3203.06\,{z}^{2}- 65.8463\,{z}^{3
}
}
{
- 0.438025- 0.334872\,z+ 0.0203996\,{z}^{2}
} 
\, Mpc
\\ 
\quad  for \quad 0.001 <z<4
\quad .
\nonumber
\label{dlz32}
\end {eqnarray}
The transversal distance in $\Lambda$CDM cosmology, 
$D_{T,3,2}$,  
which corresponds to the  angle  $\delta$ expressed in arcsec 
is  
\begin{equation}
D_{T,3,2}= \frac
{
 4.84813\,{\it \delta}\, \left(  2.328+ 502.067\,
z+ 113.03\,{z}^{2} \right)
}
{
0.124085+ 0.149501\,z+ 0.0932928\,{z}^{2}
} \, pc
\quad  .
\label{dtlcdm}
\end{equation}

\subsection{Flat Cosmology}

The two parameters of the flat cosmology 
are 
$H_0$,  the Hubble constant expressed in     $\h0units$,
and  $\om$   which is  
\begin{equation}
\om = \frac{8\pi\,G\,\rho_0}{3\,H_0^2}
\quad ,
\end{equation}
where $G$ is the Newtonian gravitational constant and
$\rho_0$ is the mass density at the present time.
In the case of $m$=2 and $n$=2 the minimax rational expression
for the luminosity distance, $\ml$, 
when $H_0=70\,\h0units$ and $\om = 0.277$,
is  
\begin{equation}
\ml =
\frac{
0.0889+ 748.555\,z+ 5.58311\,{z}^{2}
}
{
0.175804+ 0.206041\,z+ 0.068685\,{z}^{2}
}
\, 
Mpc
\quad .
\label{dlfzminimax}
\end{equation}

The transversal distance in flat cosmology, 
$D_{Tf,3,2}$,  
which corresponds to the  angle  $\delta$ expressed in arcsec 
is  
\begin{equation}
D_{Tf,3,2}= \frac
{
 4.84813\,{\it \delta}\, \left(  0.0889+ 748.555
\,z+ 5.58311\,{z}^{2} \right)
}
{
 0.175804+ 0.206041\,z+ 0.0686854\,{z}^{2}
}
\,
pc
\label{dtflat}
\end{equation}

\subsection{Modified tired light}

In an Euclidean static  framework
the modified tired light (MTL)  has been introduced in
Section 2.2 in \cite{Zaninetti2015a}.
The distance in MTL is
\begin{equation}
d= \frac{c}{H_0} \ln (1+z)
\quad .
\label{nonlzd}
\end{equation}

\begin{figure}
\begin{center}
\includegraphics[width=8cm]{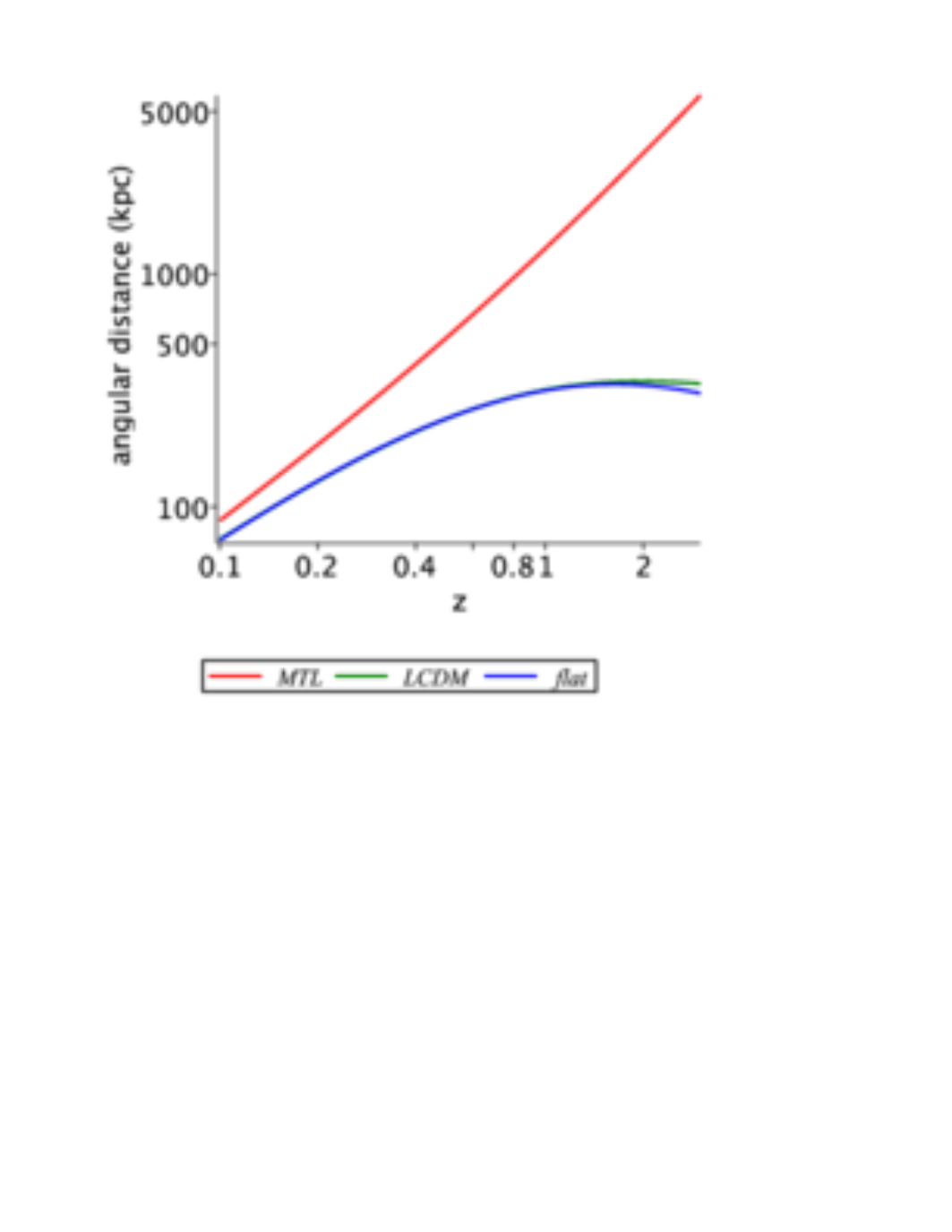}
\end{center}
\caption
{
Angular  distances in kpc 
for the three cosmologies here considered 
when $\delta=38.913456$ 
}
\label{cosmologies_angular}
\end{figure}

The  distance modulus in the modified tired light (MTL)
is
\begin{equation}
m-M = \frac{5}{2}\,{\frac {\beta\,\ln  \left( z+1 \right) }{\ln  \left( 10 \right) }
}+5\,{\frac {1}{\ln  \left( 10 \right) }\ln  \left( {\frac {\ln
 \left( z+1 \right) c}{H_{{0}}}} \right) }+25
\quad .
\label{modulustired}
\end{equation}
Here $\beta$ is a parameter comprised  between 1 and 3
which allows to match theory with observations.
The  number of free  parameters in MTL
is two: $H_0$ and $\beta$.
The fit of the distance modulus   with the data 
of Union 2.1 compilation
gives  $\beta$=2.37, $H_0=69.32 \pm 0.34 $,
see   \cite{Zaninetti2016a},
which means the following 
distance
\begin{equation}
d=4324.761\,\ln  \left( 1+z \right)  \left( 1+z \right) ^{
 1.185} \, Mpc
\quad .
\end{equation}
The transversal distance in MTL, 
$d_{T}$,  
which corresponds to the  angle  
$\delta$ expressed in arcsec 
is  
\begin{equation}
d_T= 
20967\,{\it \delta }\,\ln  \left( 1+z \right)  \left( 1
+z \right) ^{ 1.185}
\, pc
\label{dtmtl}
\end{equation}
We report  the angular distance for a fixed
$delta$ as function of redshift for the three 
cosmologies, see Figure \ref{cosmologies_angular}.
The angular distance in flat and $\Lambda$CDM cosmology
does not increase with z, see \cite{Peebles1993},
in contrast with   
the modified tired light .

\section{The motion of a SB}

\label{sec_sb}
We now summarize the  adopted profile of
density and the equation of motion
for a SB.

\subsection{The profile}

The density  is  assumed to have the following hyperbolic 
dependence on $Z$ which is the third 
 Cartesian coordinate,
\begin{equation}
 \rho(Z;Z_0,\rho_0) =
   \left \{
    \begin{array}{ll}
    \rho_0                 & \mbox {if  z   $\leq Z_0$ } \\
    \rho_0 \frac{Z_0}{z}   & \mbox {if  z   $>    Z_0$ } \\
  \end{array}
  \label{profhyperbolic}
\right.
\end{equation}
where the parameter $Z_0$ fixes the scale and  $\rho_0$ is the
density at $Z=Z_0$.
In spherical coordinates
the dependence  on the polar angle is
\begin{equation}
 \rho(r;\theta,Z_0,\rho_0) =
   \left\{
    \begin{array}{ll}
    \rho_0                      
    & \mbox{ if $\cos(\theta)  \leq Z_0$} \\
    \rho_0 \frac{Z_0}{r \cos(\theta)}  
    & \mbox{if  $r \cos(\theta)  >    Z_0$}\\
  \end{array}
  \right. 
 \label{profhyperbolicr}
\end{equation}
Given a solid angle  $\Delta \Omega$
the mass $M_0$ swept
in the interval $[0,r_0]$
is
\begin{equation}
M_0 =
\frac{1}{3}\,\rho_{{0}}\,{r_{{0}}}^{3} \Delta \Omega
\quad .
\end{equation}
The total mass swept, 
$M(r;r_0,Z_0,\alpha,\theta,\rho_0) $,
in the interval $[0,r]$
is
\begin{equation}
M(r;r_0,Z_0,\alpha,\theta,\rho_0)= \bigl (
\frac{1}{3}\,\rho_{{0}}{r_{{0}}}^{3}+\frac{1}{2}\,{\frac {\rho_{{0}}Z_{{0}} \left( {r}
^{2}-{r_{{0}}}^{2} \right) }{\cos \left( \theta \right) }}
\bigr) 
\Delta \Omega
\quad .
\label{masshyperbolic}
\end{equation}
and its  approximate value   at high values of $r$
is  
\begin{equation}
M(r;Z_0,\alpha,\theta,\rho_0) \approx 
\frac{1}{2}\,{\frac {{r}^{2}\rho_{{0}}Z_{{0}}}{\cos \left( \theta \right) }}
\Delta \Omega
\quad .
\label{approximatemass}
\end{equation}
The density $\rho_0$ can be  obtained
by introducing  the number density, $n_0$, expressed  in particles
$\mathrm{cm}^{-3}$,
the mass of  hydrogen, $m_H$,
and  a multiplicative factor $f$,
which is chosen to be  1.4, see \cite{McCray1987},
\begin{equation}
\rho_0  = f  m_H n_0
\quad .
\end{equation}
The astrophysical version of the total approximate swept mass
as  given by equation (\ref{approximatemass}),
expressed in solar mass
units, $M_{\sun}$,  is 
\begin{equation}
M (r_{pc};Z_{0,pc},n_0,\theta)\approx
\frac{0.0172 \,n_{{0}}{   z}_{{{   0,pc}}}{r_{{{   pc}}}}^{2}}
{cos(\theta)}
\,M_{\sun}\,\Delta \Omega 
\quad ,
\end{equation}
where
$Z_{0,pc}$,  and $r_{0,pc}$
are  $Z_0$,  and $r$ expressed  in pc.
\subsection{The equation of motion}

The conservation of the classical momentum in
spherical coordinates
along  the  solid angle  $\Delta \Omega$
in the framework of the thin
layer approximation  states that
\begin{equation}
M_0(r_0) \,v_0 = M(r)\,v
\quad ,
\end{equation}
where $M_0(r_0)$ and $M(r)$ are the swept masses at $r_0$ and $r$,
and $v_0$ and $v$ are the velocities of the thin layer at $r_0$ and $r$.
This conservation law can be expressed as a differential equation
of the first order by inserting $v=\frac{dr}{dt}$:
\begin{equation}
M(r)\, \frac{dr}{dt} - M_0\, v_0=0
\quad .
\end{equation}
The velocity  as a function of the radius $r$  is
\begin{equation}
v(r;r_0,Z_0,v_0,\theta)=
2\,{\frac {{r_{{0}}}^{3}v_{{0}}\cos \left( \theta \right) }{2\,{r_{{0}
}}^{3}\cos \left( \theta \right) -3\,{r_{{0}}}^{2}Z_{{0}}+3\,{r}^{2}Z_
{{0}}}}
\label{vranalyticalhyper}
\quad .
\end{equation}
The differential equation
which models the momentum conservation
in the case of a  hyperbolic profile is
\begin{equation}
 \left( \frac{1}{3}\, {r_{{0}}}^{3}+\frac{1}{2}\,{\frac { Z_{{0}}
 \left( -{r_{{0}}}^{2}+ \left( r \left( t \right)  \right) ^{2}
 \right) }{\cos \left( \theta \right) }} \right) {\frac {\rm d}{
{\rm d}t}}r \left( t \right) -\frac{1}{3}\, {r_{{0}}}^{3}v_{{0}}=0
\quad ,
\end{equation}
where the initial  conditions
are  $r=r_0$  and   $v=v_0$
when $t=t_0$.

The variables can be separated and
the radius as a function of the time
is
\begin{equation}
r(t;t_0,r_0,Z_0,v_0,\theta) = \frac{HN}{HD}
\quad ,
\nonumber
\label{rtanalyticalhyper}
\end{equation}
where
\begin{eqnarray}
HN= 
-\sqrt [3]{3} \Bigg  ( 2\,\cos   ( \theta   ) \sqrt [3]{3}r_{{0}}
-3\,\sqrt [3]{3}Z_{{0}}- \bigg  ( -9\,{Z_{{0}}}^{3/2}+   (    ( 9
\,t-9\,t_{{0}}   ) v_{{0}}
\nonumber  \\
+9\,r_{{0}}   ) \cos   ( \theta
   ) \sqrt {Z_{{0}}}+\sqrt {3}\sqrt {27}\sqrt {{\it AHN}} \bigg  ) 
^{2/3}  \Bigg ) r_{{0}}
\quad ,
\end{eqnarray}
with

\begin{eqnarray}
AHN= 
 \Bigg ( {\frac {8\,   ( \cos   ( \theta   )    ) ^{2}{r_
{{0}}}^{3}}{27}}
\nonumber \\
+Z_{{0}}  \bigg (    ( t-t_{{0}}   ) ^{2}{v_{{0}}
}^{2}+2\,r_{{0}}   ( t-t_{{0}}   ) v_{{0}}-\frac{1}{3}\,{r_{{0}}}^{2}
  \bigg ) \cos   ( \theta   )
\nonumber \\
 -2\,v_{{0}}{Z_{{0}}}^{2}   ( t-
t_{{0}}   )  \Bigg  ) \cos   ( \theta   ) 
\end{eqnarray}
and
\begin{eqnarray}
HD = 
3\,\sqrt {Z_{{0}}} \times \nonumber\\ \sqrt [3]{-9\,{Z_{{0}}}^{3/2}+ \left(  \left( 9\,t-9
\,t_{{0}} \right) v_{{0}}+9\,r_{{0}} \right) \cos \left( \theta
 \right) \sqrt {Z_{{0}}}+9\,\sqrt {{\it BHD}}}
\quad ,  
\end{eqnarray}
with   
\begin{eqnarray}
BHD=
\Bigg   ( {\frac {8\,   ( \cos   ( \theta   )    ) ^{2}{r_
{{0}}}^{3}}{27}}+ \nonumber \\
Z_{{0}} \bigg  (    ( t-t_{{0}}   ) ^{2}{v_{{0}}
}^{2}+2\,r_{{0}}   ( t-t_{{0}}   ) v_{{0}}-\frac{1}{3}\,{r_{{0}}}^{2}
\bigg   ) \cos   ( \theta   )
\nonumber  \\
 -2\,v_{{0}}{Z_{{0}}}^{2}   ( t-
t_{{0}}   )   \Bigg  ) \cos   ( \theta   ) 
\quad .
\end{eqnarray}
As a consequence  the velocity as function
of the time is  
\begin{equation}
v(t;t_0,r_0,Z_0,v_0,\theta) = \frac{d r(t;t_0,r_0,Z_0,v_0,\theta)}{dt}
\quad .
\nonumber
\label{vtanalyticalhyper}
\end{equation}
More details as well the exploration of other 
profiles of density can be found in \cite{Zaninetti2018a}.
We now continue evaluating the flux of kinetic energy, $F_{ek}$,
in the thin emitting layer which is supposed to have
density $\rho_l$ 
\begin{equation}
F_{ek}(t;t_0,r_0,Z_0,v_0,\theta) = \frac{1}{2} \rho_l 4 \pi r(t)^2 v(t)^3
\quad .
\label{fluxek}
\end{equation}
The volume  of the  thin emitting layer, $V_l$,
is approximated by 
\begin{equation}
V_l = 4\,{\it \Delta}\,\pi\,{r}^{2}
\quad ,
\label{approximatevolume}
\end{equation}
where $\Delta$ is thickness   of the layer;
as  an example   \cite{McCray1987} quotes  $\Delta =\frac{r}{12}$.
The two  approximations for mass, equation~(\ref{approximatemass}),
and  volume,  equation~(\ref{approximatevolume}),  
allows to derive an approximate value for the density 
in the thin layer 
\begin{equation}
\rho_l= \frac{1}{8}\,{\frac {\rho_{{0}}Z_{{0}}f}{\cos \left( \theta \right) r\pi}}
\label{approximatedensity}
\quad .
\end{equation}
Inserted in   equation  (\ref{fluxek}) the radius, velocity 
and density 
as given by 
equations (\ref{rtanalyticalhyper}),
 (\ref{vtanalyticalhyper})
and (\ref{approximatedensity}),
we obtain 
\begin{equation}
F_{ek}(t;t_0,r_0,Z_0,v_0,\theta) =\frac{FN}{FD}
\quad ,
\label{fektheo}
\end{equation}
where 
\begin{eqnarray}
FN=\quad  ,
-\sqrt {27} \left( -3\,\sqrt {3}{Z_{{0}}}^{3/2}+\sqrt {27}\sqrt {F_{{1
}}\cos \left( \theta \right) }+3\,F_{{5}} \right) ^{3}f 
\times 
\nonumber  \\
\left( 2\,
\sqrt [3]{3}F_{{3}}+ \left( -9\,{Z_{{0}}}^{3/2}+F_{{2}}+\sqrt {27}
\sqrt {F_{{1}}\cos \left( \theta \right) }\sqrt {3} \right) ^{2/3}
 \right) ^{3}{r_{{0}}}^{4}
\times  
\nonumber  \\
\cos \left( \theta \right) 
{{\it v0}}^{3}
\sqrt {Z_{{0}}}\bigg   ( 2\,\sqrt [3]{3}F_{{3}}- \big  ( -9\,{Z_{{0}}}^{3
/2}+F_{{2}}+
\nonumber  \\
\sqrt {27}\sqrt {F_{{1}}\cos   ( \theta   ) }\sqrt 
{3} \big  ) ^{2/3} \bigg  ) \sqrt [3]{3}\rho_{{0}}
\end{eqnarray}
and  
\begin{eqnarray}
FD=
108\,\sqrt {F_{{1}}\cos \left( \theta \right) }
\times  \nonumber\\ 
 \left( -9\,{Z_{{0}}}^{
3/2}+F_{{2}}+\sqrt {27}\sqrt {F_{{1}}\cos \left( \theta \right) }
\sqrt {3} \right) ^{13/3}F_{{4}}
\end{eqnarray}
being   
\begin{eqnarray}
F_1 =
{\frac {8  \left( \cos \left( \theta \right)  \right) ^{2}{r_{{0}}}^{
3}}{27}}+ \left(  \left( t-t_{{0}} \right) ^{2}{{\it v0}}^{2}+2 r_{{0
}} \left( t-t_{{0}} \right) {\it v0}-\frac{1}{3} {r_{{0}}}^{2} \right) \times
\nonumber \\
 Z_{{0}
}\cos \left( \theta \right) -2 {\it v0} {Z_{{0}}}^{2} \left( t-t_{{0
}} \right)
\quad  ,
\end{eqnarray}
\begin{eqnarray}
F_2 =\left(  \left( 9 t-9 t_{{0}} \right) {\it v0}+9 r_{{0}} \right) 
\cos \left( \theta \right) \sqrt {Z_{{0}}}
\quad  ,
\end{eqnarray}
\begin{eqnarray}
F_3 =
\cos \left( \theta \right) r_{{0}}-3/2 Z_{{0}}
\end{eqnarray}
\begin{eqnarray}
F_4 =
8\,   ( \cos   ( \theta   )    ) ^{2}{r_{{0}}}^{3}+27\,
   \Bigr (    ( t-t_{{0}}   ) ^{2}{{\it v0}}^{2}+2\,r_{{0}}
   ( t-t_{{0}}   ) {\it v0}-
\nonumber \\
\frac{1}{3}\,{r_{{0}}}^{2}  \Bigl ) Z_{{0}}
\cos   ( \theta   ) -54\,{\it v0}\,{Z_{{0}}}^{2}   ( t-t_{{0
}}   ) 
\quad  ,
\end{eqnarray}
\begin{eqnarray}
F_5 =
\left( {\it v0}\, \left( t-t_{{0}} \right) +r_{{0}} \right) \cos
 \left( \theta \right) \sqrt {3}\sqrt {Z_{{0}}}
\quad  .
\end{eqnarray}

We now assumes  that the amount of luminosity, $L_{theo}$,
reversed in the shocked emission is proportional 
to the flux of kinetic energy as given by   
equation (\ref{fektheo}) 
\begin{equation}
L_{theo} (t;t_0,r_0,Z_0,v_0,\theta) 
\propto 
F_{ek}(t;t_0,r_0,Z_0,v_0,\theta)
\quad .
\end{equation}
The theoretical luminosity  is  not equal  along all the SB 
but is function of the polar angle $\theta$.
In this framework is useful to introduce the ratio, $\kappa$,
between
theoretical luminosity at $\theta$ and 
and that one at 
$\theta=0$,
\begin{equation}
\kappa =
\frac
{
L_{theo} (t;t_0,r_0,Z_0,v_0,\theta) 
}
{ 
L_{theo}(t;t_0,r_0,Z_0,v_0,\theta=0)
}
\quad .
\label{ratiok}
\end{equation}

The above model for the theoretical luminosity 
is independent from the image theory, see Section~\ref{imagetheory},
and  does not explains the hole of luminosity 
visible  in the  shells.

\section{Astrophysical Environment}

\label{sec_astro}
We now analyze  
the catalogue for the giant arcs,
the two giant arcs SDP.81 and A2267
and the initial astrophysical conditions
for the SBs.

\subsection{The catalogue}

\begin{figure}
\begin{center}
\includegraphics[width=8cm,angle=-90]{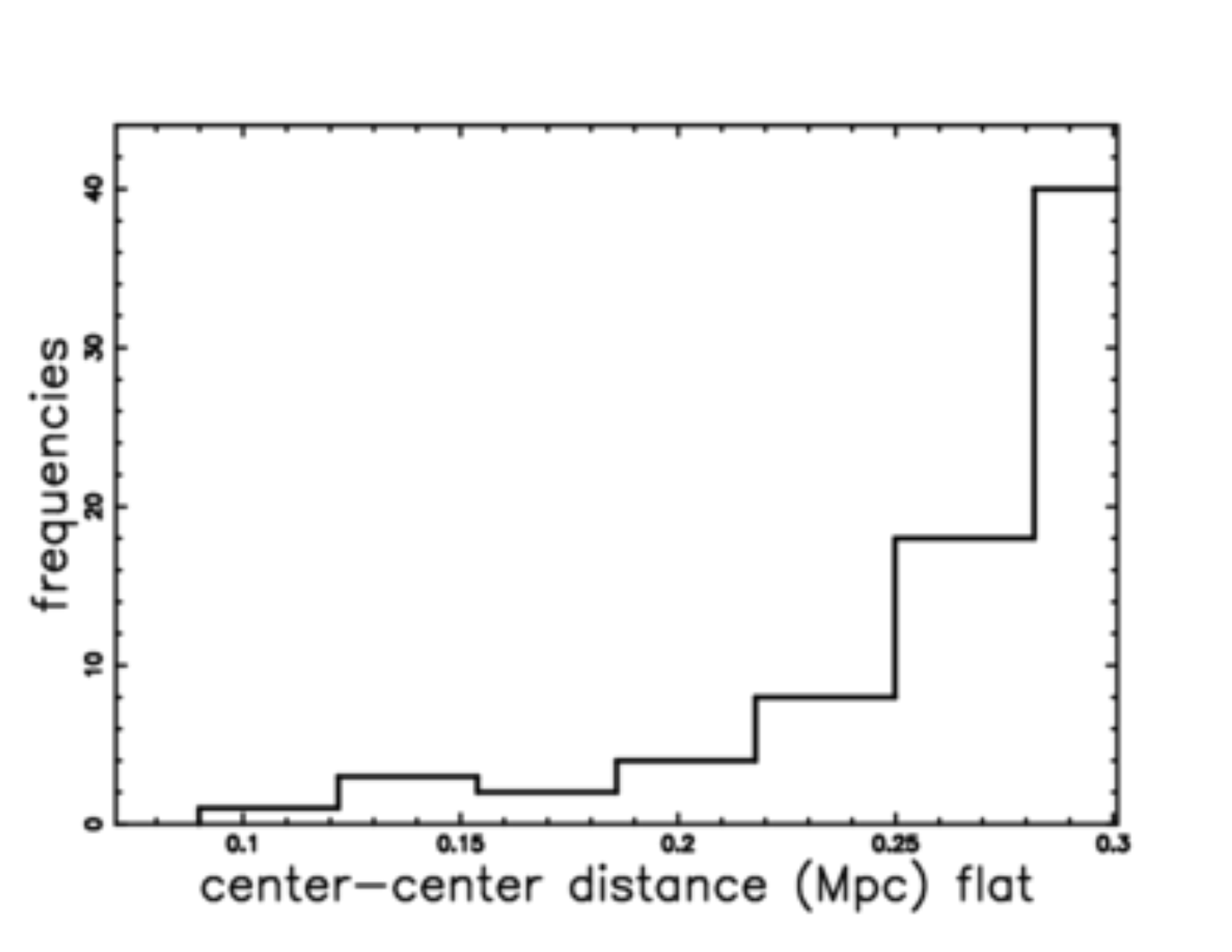}
\end{center}
\caption
{
Histogram of the radial distance from the arc center 
to the cluster center in Mpc  
in flat cosmology with conversion formula (\ref{dtflat}).
}
\label{center_center_flat}
\end{figure}

\begin{figure}
\begin{center}
\includegraphics[width=8cm,angle=-90]{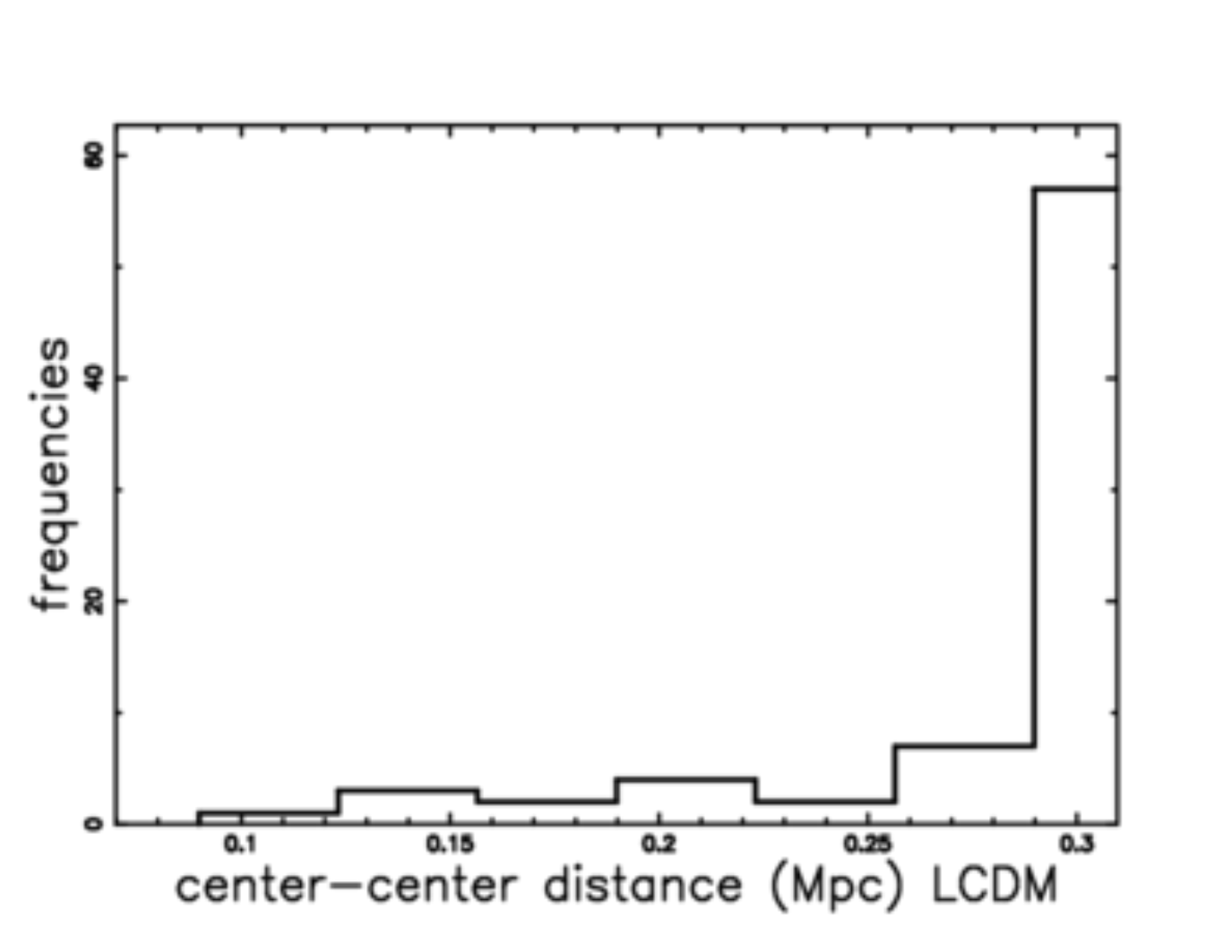}
\end{center}
\caption
{
Histogram of the radial distance from the arc center 
to the cluster center in Mpc  
in $\Lambda$CDM cosmology with conversion formula (\ref{dtlcdm}).
}
\label{center_center_lcdm}
\end{figure}

\begin{figure}
\begin{center}
\includegraphics[width=8cm,angle=-90]{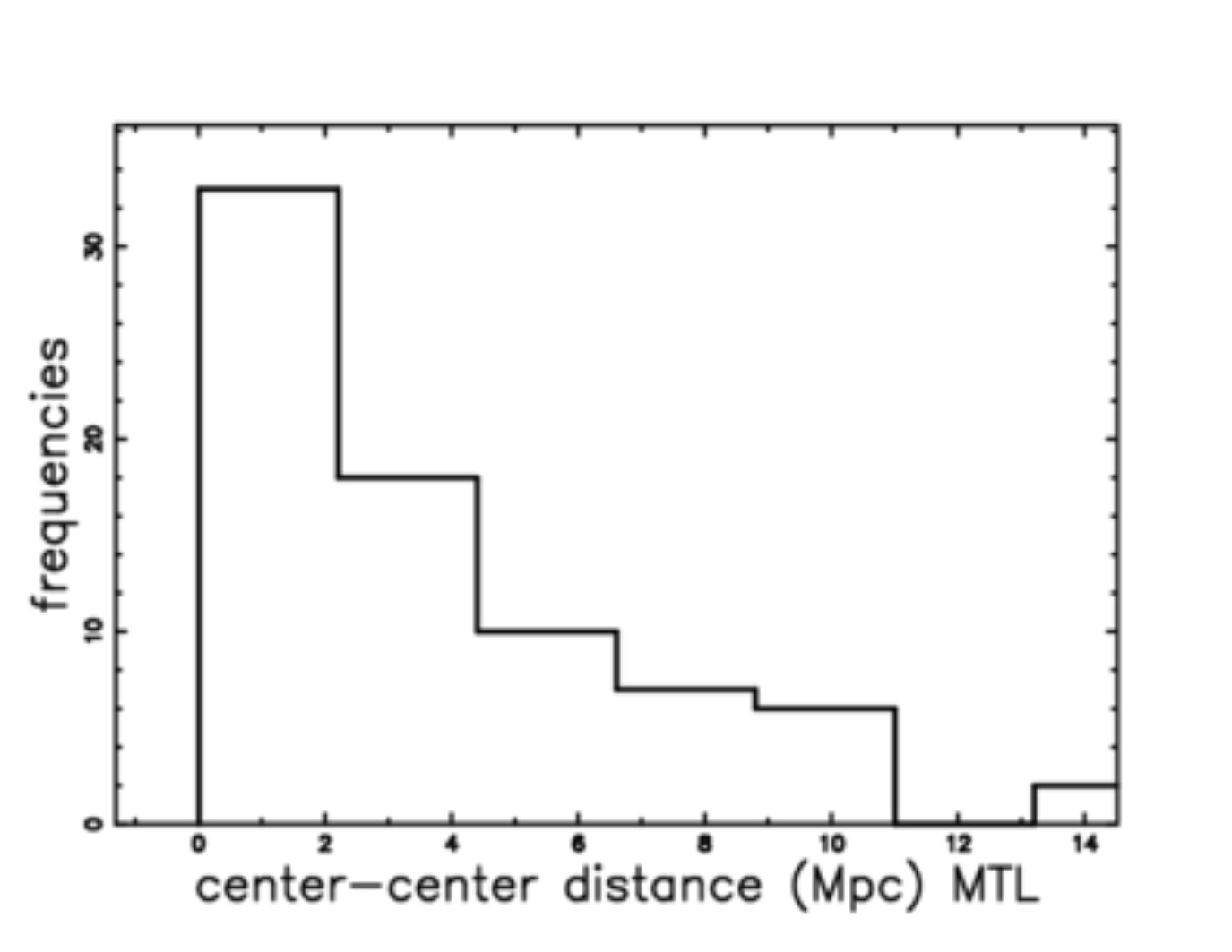}
\end{center}
\caption
{
Histogram of the radial distance from the arc center 
to the cluster center in Mpc  
in MTL cosmology with conversion formula (\ref{dtmtl}).
}
\label{center_center_mtl}
\end{figure}

Some  parameters   of the giant arcs as detected  as images  by 
cluster lensing and the 
supernova survey with Hubble (CLASH)
which is  available as   a catalogue  
at  \url{http://vizier.u-strasbg.fr/viz-bin/VizieR},
see \cite{Xu2016}.
We are interested in 
the arc length which is given in arcsec,
the arc length to width ratio,
the  photometric redshift,
and in
the radial distance from the arc center to the cluster center in arcsec.
Table \ref{center_center} reports the  
statistical parameters  of the
radial distance from the arc center to the cluster center in kpc
and 
Figure \ref{center_center_flat},
Figure \ref{center_center_lcdm},
and Figure \ref{center_center_mtl} 
the histogram of the frequencies 
in the framework of flat,$\Lambda$CDM and
MTL cosmology respectively.
\begin{table}[ht!]
\caption
{
Statistical parameters of the
radial distance from the arc center 
to the cluster center in kpc  
}
\label{center_center}
\begin{center}
{
\begin{tabular}{|c|c|c|c|}
\hline
Cosmology  &  minimum (kpc) & average (kpc)    &  maximum (Mpc)     \\
\hline
Flat         & 89 &  270  & 313    \\
$\Lambda$CDM & 89 &  288  & 323    \\
MTL          & 7  &  3810 & 1540   \\
\hline
\end{tabular}
}
\end{center}
\end{table}

\subsection{Single giant arcs}

The ring associated with the galaxy SDP.81,
see \cite{Eales2010}, 
is characterized by a foreground galaxy at 
$z=0.2999$ and a background galaxy at $z=0.3042$.
This ring  has  been studied with the 
Atacama Large Millimeter/sub-millimeter Array (ALMA)
by 
\cite{Tamura2015,ALMA2015,Rybak2015,Hatsukade2015,Wong2015,Hezaveh2016}
and  has the 
observed parameters as in Table \ref{paragiants}.
\begin{table}[ht!]
\label{paragiants}
\caption
{
Observed  parameters of the giants arcs.
}
\label{}
\begin{center}
{
\begin{tabular}{|c|c|c|}
\hline
Name   & redshift  & radius~arcsec \\
\hline  
SDP.81 & 3.04      & 1.54          \\
A2667  & 1.033     & 42            \\
\hline
\end{tabular}
}
\end{center}
\end{table}

Another  giant arc is that in A2667  which is made by three pieces:
A, B and C , see Figure 1 in  \cite{Yuan2012}.
The  radius can be found  from the equation of  the circle 
given the three points
A, B and C, see Table \ref{radiuskpc}.
The three pieces can be digitalized for a further 
comparison with a simulation,
see  empty red stars in Figure \ref{a2267_real}.

\begin{table}[ht!]
\caption
{
Radius of the giant arcs in kpc.
}
\label{radiuskpc}
\begin{center}
{
\begin{tabular}{|c|c|c|}
\hline
Cosmology   &  SDP.81 & A2667     \\
\hline
Flat         & 12.09 & 345.39   \\
$\Lambda$CDM & 13.33 & 347.42  \\
MTL          & 235.82 &  1456.32   \\
\hline
\end{tabular}
}
\end{center}
\end{table}

\subsection{The initial conditions}

We review the  starting equations for the evolution
of the SB \cite{Dyson1997,mccrayapj87,Zaninetti2004}
which can be derived from  the   momentum conservation
 applied to a pyramidal section.
The  parameters of the thermal model are 
$N^*$,
the number of SN explosions in  $5.0 \cdot 10^7$ \mbox{yr},
$Z_{\mathrm{OB}}$,
the distance of the OB associations from the galactic plane,
$E_{51}$, 
the  energy in  $10^{51}$ \mbox{erg} usually chosen equal to one,
$v_0$, 
the initial velocity which is fixed by the bursting phase,
$t_0$,
the initial time in $yr$  which is equal to the bursting time,
and $t$  the proper time  of the SB.
With the above  definitions the  radius of the SB 
is
\begin{equation}
R =111.56\,(\frac{E_{51}t_7^3 N^*}{n_0})^{\frac {1} {5}}
\,\mathrm{pc},
\label{raggioburst}
\end{equation}
and its velocity 
\begin{equation}
V= 6.567\,{\frac {1}{{{\it t_7}}^{2/5}}\sqrt [5]{{\frac {E_{{51}}
{\it N^*}}{n_{{0}}}}}} \,\mathrm{ \frac{km}{s}} 
\quad .
\end{equation}
In the following, we will  assume that 
the bursting phase  ends at $t=t_{7,0}$   (the bursting time is expressed in
units of $10^7$ yr) 
when  $N_{SN}$ SN are exploded 
\begin{equation}
N_{SN} = N^* \frac{t_{7,0} \cdot  10^7} {5 \cdot 10^7}
\quad .
\end{equation}   
The two following  inverted formula allows to derive  
the parameters of the initial conditions for the SB
with ours $r_0$ expressed in pc  and $v_0$ expressed 
in $km\,s^{-1}$  are
\begin{equation}
t_{7,0}= 0.05878095238\,{\frac {r_{{0}}}{v_{{0}}}} 
\quad ,
\end{equation}
and
\begin{equation}
N^*= 2.8289\,10^{-7}\,{\frac {{r_{{0}}}^{2}n_{{0}}{v_{{0}}}^{3}}{E_{{51
}}}}
\quad  .
\end{equation}

\section{Astrophysical Simulation}

\label{sec_simulation}
We simulate a single giant arc, A2267,
and then we simulate the statistics
of many giant arcs.

\subsection{Simulation of A2667}

\begin{figure}
\begin{center}
\includegraphics[width=8cm]{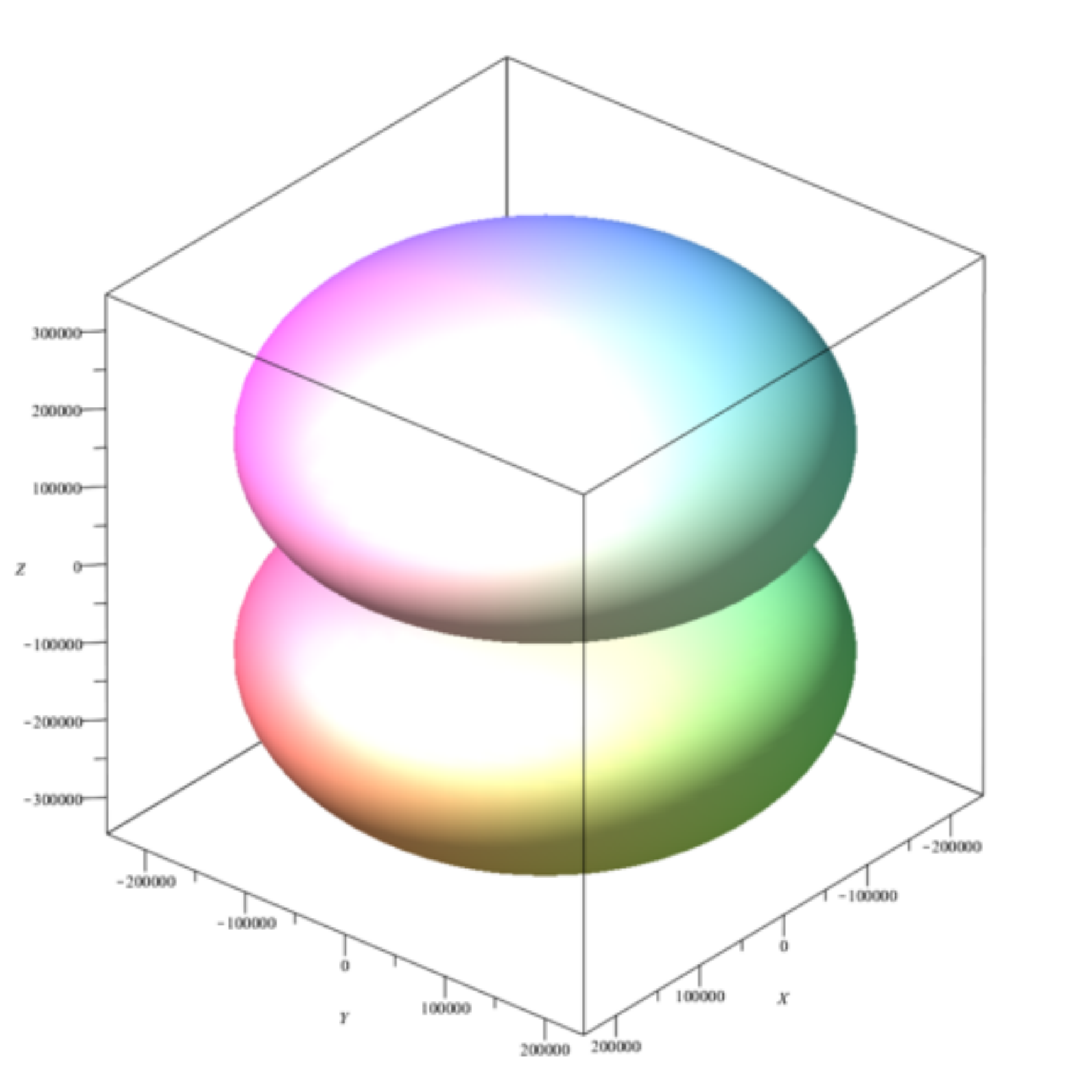}
\end{center}
\caption
{
3D surface  of the SB connected with A2267,
parameters as in Table \ref{parasba2267}
and axes in pc.
}
\label{3dsurfacegiantsarcs}
\end{figure}

\begin{figure}
\begin{center}
\includegraphics[width=8cm,angle=-90]{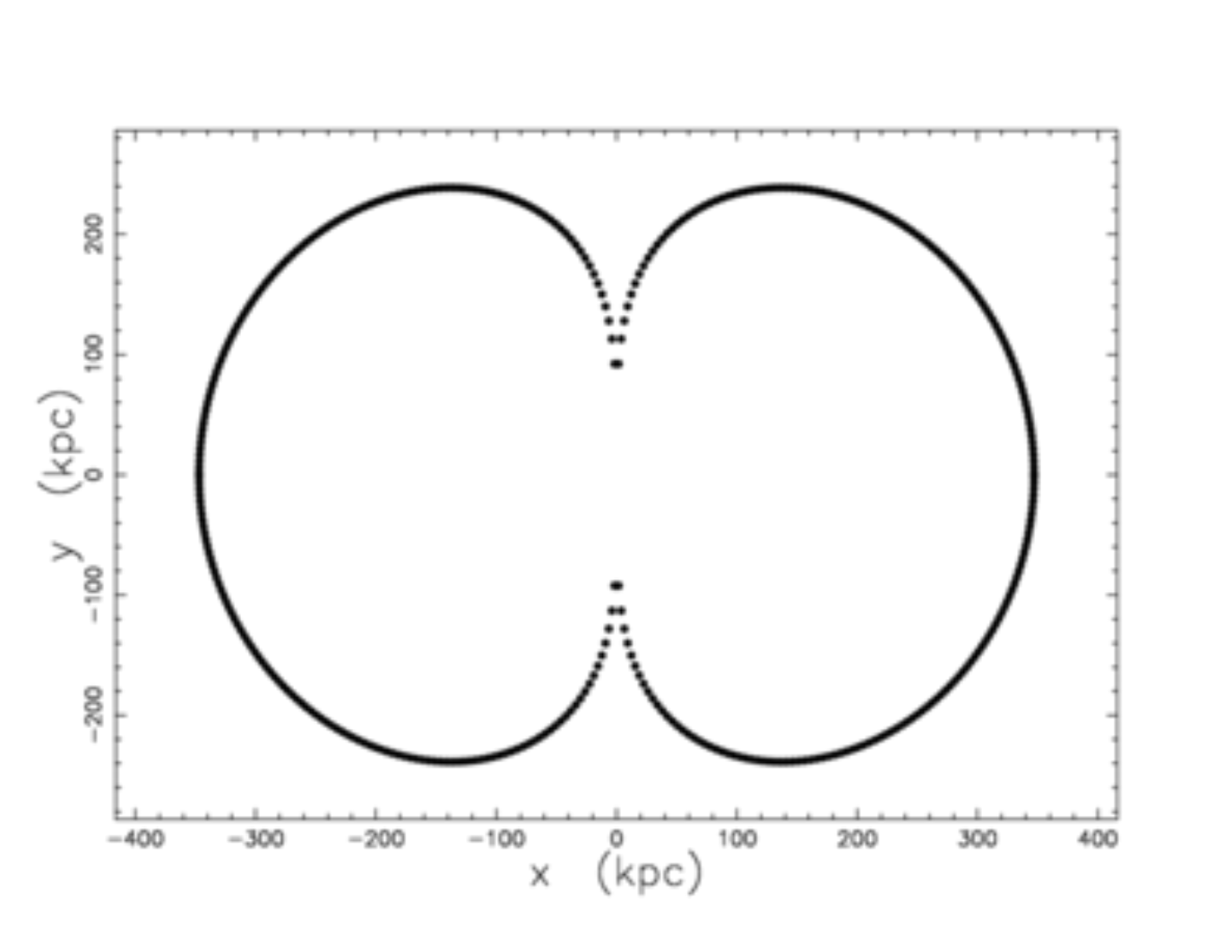}
\end{center}
\caption
{
2D section in the $z=0$ plane   of the SB connected with A2267,
parameters as in Table \ref{parasba2267}
and axes in kpc.
}
\label{a2267_theo}
\end{figure}

The final stage   of the SB connected with A2267
is   simulated with the parameters reported 
in Table \ref{parasba2267}~;
in particular 
Figure \ref{3dsurfacegiantsarcs} displays 
the 3D shape and 
Figure~\ref{a2267_theo} reports the 2D section.

\begin{table}[ht!]

\caption
{
Theoretical parameters of the SB connected with A2267.
}
\label{parasba2267}
\begin{center}
{
\begin{tabular}{|c|c|c|}
\hline
theory  & parameter   & value \\
\hline  
initial~thermal~model & $E_{51}$       & 1                \\
initial~thermal~model & $n_0$          & 1                \\
initial~thermal~model & $t_{7,0}$      &  0.0078          \\
initial~thermal~model & $N^*$          &  $1.22\,10^{14}$ \\
initial~thermal~model & $N_{SN}$       &  $1.91\,10^{11}$ \\
\hline
SB                    & $r_0$          &  4000~pc         \\
SB                    & $Z_0$          &  74.07~pc       \\
SB                    & $v_0$          &  30000~km/s       \\
SB                    & t              &  $8\,10^8$~yr  \\
\hline
\end{tabular}
}
\end{center}
\end{table}

\begin{figure}
\begin{center}
\includegraphics[width=8cm,angle=-90]{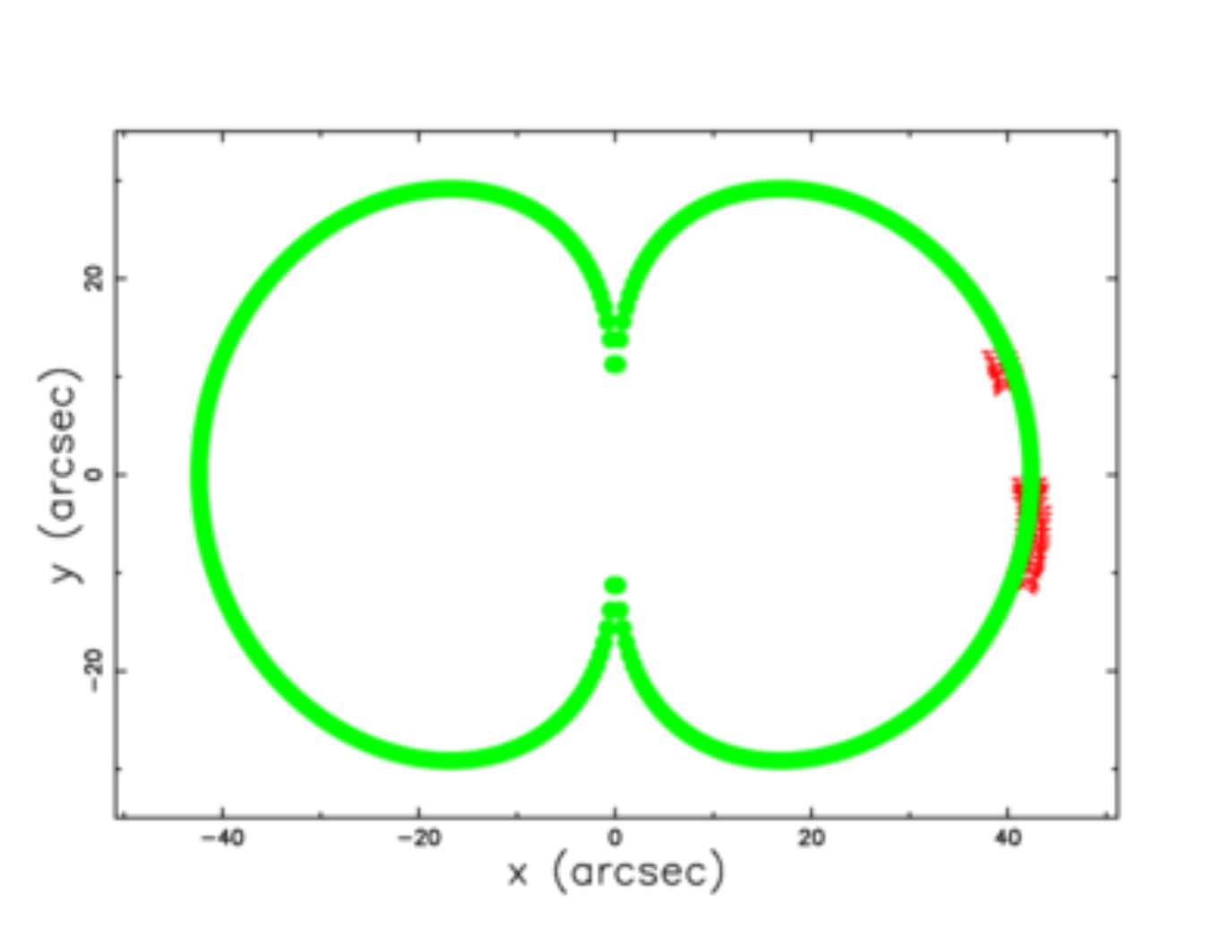}
\end{center}
\caption
{
2D section in the $z=0$ plane of the SB connected with A2267,
parameters as in Table \ref{parasba2267}
(full green points)
and
the three pieces of the giant arc in A2667 (empty red stars);
axes in arcsec.
}
\label{a2267_arcsec}
\end{figure}

\begin{figure}
\begin{center}
\includegraphics[width=8cm,angle=-90]{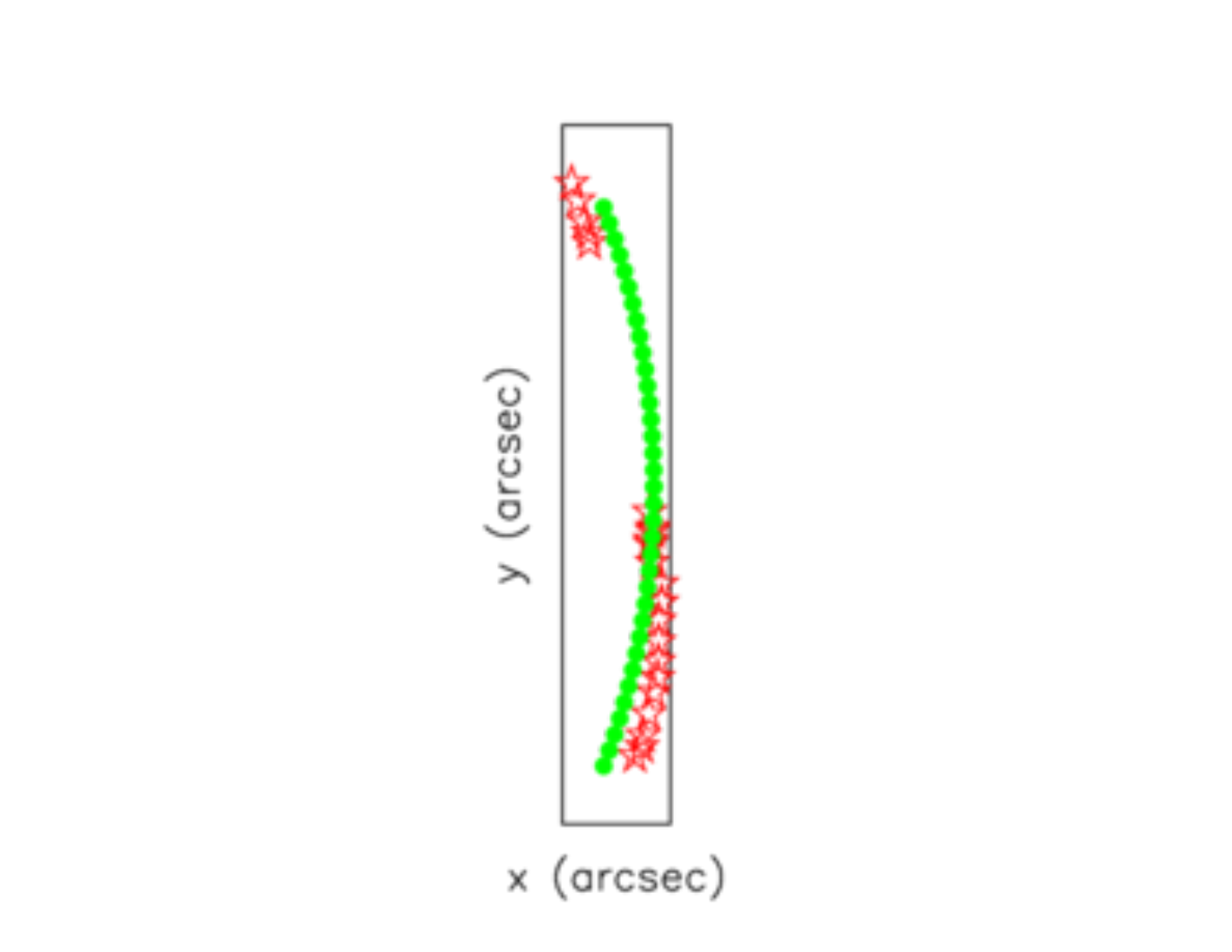}
\end{center}
\caption
{
Enlarged view of the three pieces of the giant arc in A2667 
(empty red stars)
and the theoretical radius (full green points);
axes in arcsec.
}
\label{a2267_real}
\end{figure}

Figure  \ref{a2267_arcsec} reports  the  
2D section  of the SB as well  the three pieces 
of the giant arc connected with A2267.
The similarity  between the 
observed radius of curvature of the giant arc as well
the theoretical one is reported in a zoom, 
see Figure~\ref{a2267_real}.

\begin{figure}
\begin{center}
\includegraphics[width=8cm,angle=-90]{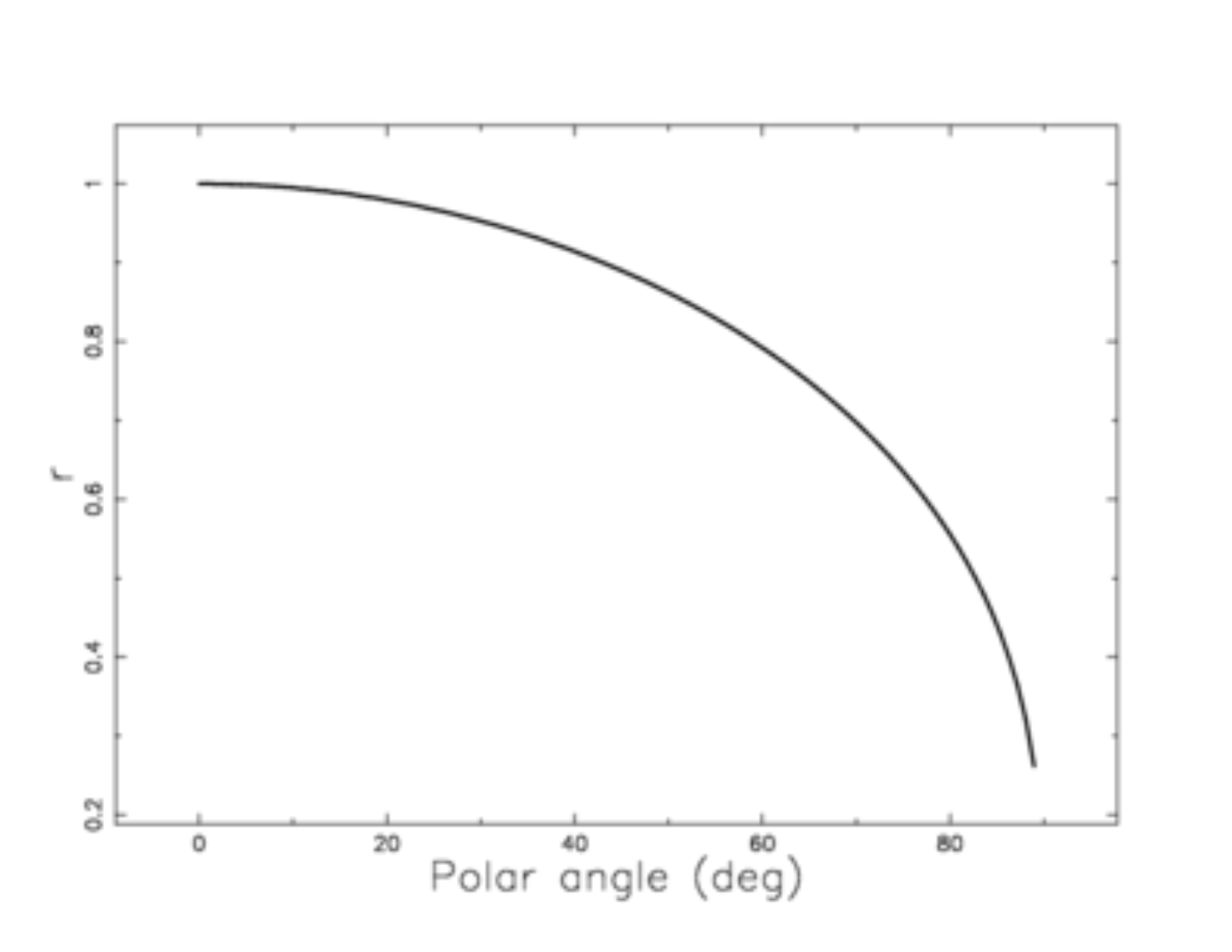}
\end{center}
\caption
{
Normalized luminosity as function 
of the polar angle in deg,
parameters as in Table \ref{parasba2267}
}
\label{a2267_rapp}
\end{figure}

We can understand  the reason for which the giant arc A2267 
has a limited angular extension of $\approx~31^{\circ}$ 
by plotting the ratio $\kappa$, equation (\ref{ratiok}),
between the theoretical luminosity as function of $\theta$ 
and the  theoretical luminosity at $\theta=0$
with parameters as in Table \ref{parasba2267},
see Figure \ref{a2267_rapp}.
As a practical example  at  $\approx~31^{\circ}/2 $, where the factor two
arises from the symmetry of the framework, 
the theoretical luminosity is decreased of a factor $\kappa= 0.987$ 
in respect to the value at $\theta=0$.
We now  introduce the threshold luminosity, $L_{tr}$,
which is an observational parameter.
The theoretical luminosity will scale as function of
the polar angle as $L_{theo}(\theta) \propto L_0 *r$ 
where $L_0$ is the theoretical luminosity  at $\theta=0$
and $\kappa$ has been defined in equation~(\ref{ratiok}). 
When the inequality $L_{theo}<L_{tr}$ is verified 
the giant arc is impossible to detect 
and only the zone characterized by low values of the polar
angle will be detected.

\begin{figure}
\begin{center}
\includegraphics[width=8cm,angle=-90]{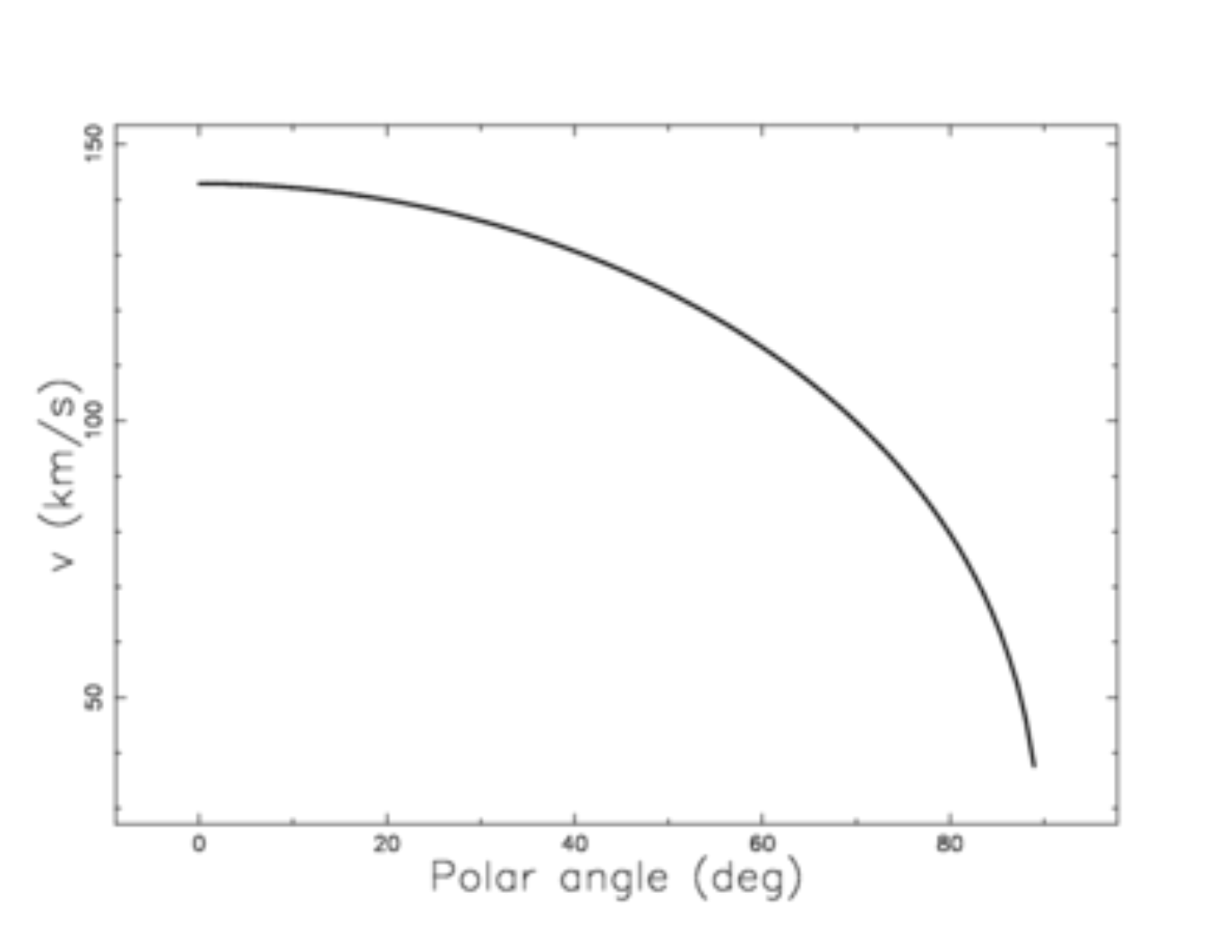}
\end{center}
\caption
{
Velocity in km/s as   function 
of the polar angle in deg.
}
\label{a2267_vel}
\end{figure}

In our model the velocity 
with parameters as in Table \ref{parasba2267}
is function of  
the polar angle, see Figure~\ref{a2267_vel},
and has range $37~km/s\,<\,v(\theta)\,<\,142\,km/s$. 
As a comparison  a velocity 
$50~km/s\,<\,v\,<\,75\,km/s$ is measured
in A2267, see Figure 5 in  \cite{Yuan2012}. 
 
\subsection{Simulation of many giants arcs}

\begin{figure}
\begin{center}
\includegraphics[width=8cm,angle=-90]{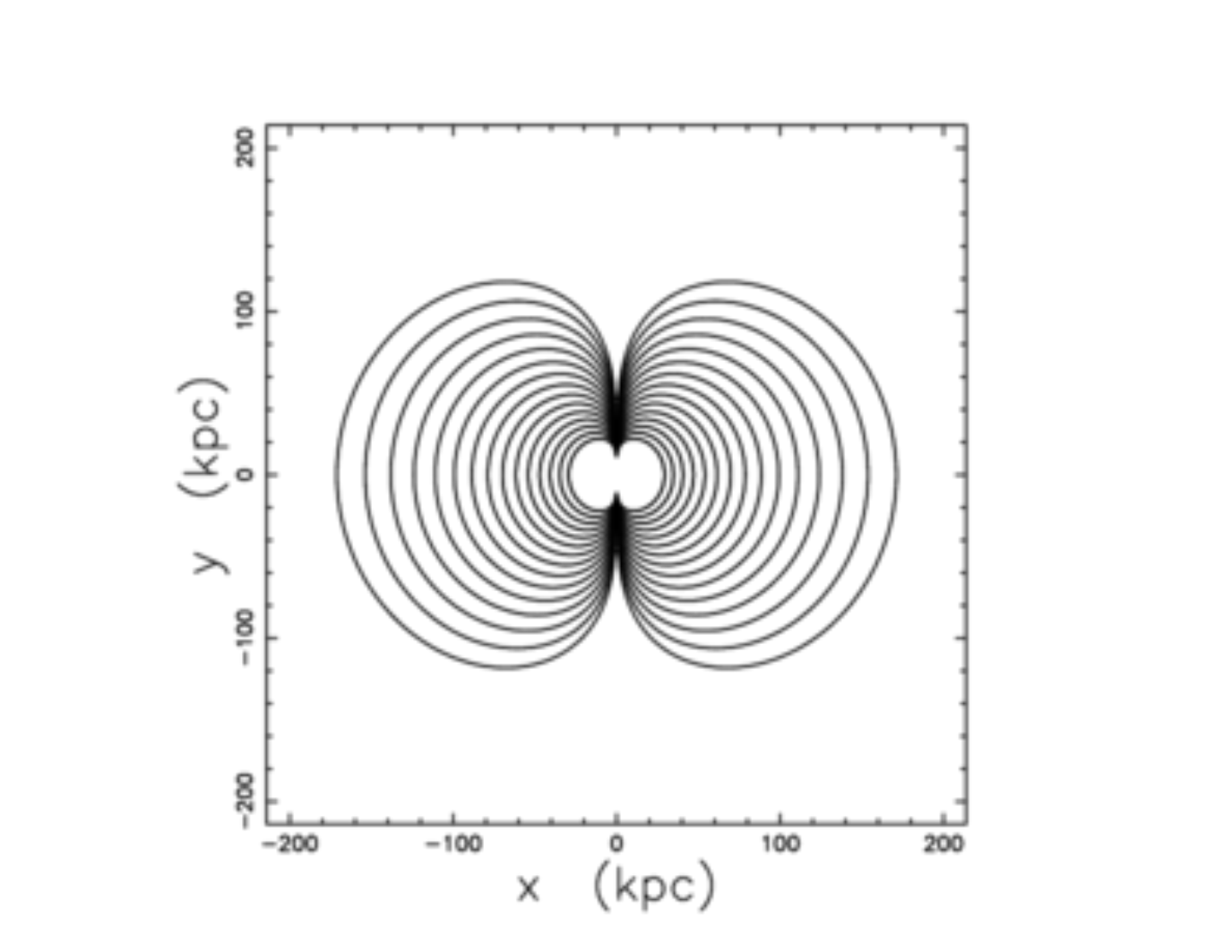}
\end{center}
\caption
{
Multiple sections of the SB with time,$t$ comprised in  
$[10^6~yr,~10^8~yr]$  
and other parameters as in Table \ref{parasba2267}.
}
\label{allsb}
\end{figure}

\begin{figure}
\begin{center}
\includegraphics[width=8cm,angle=-90]{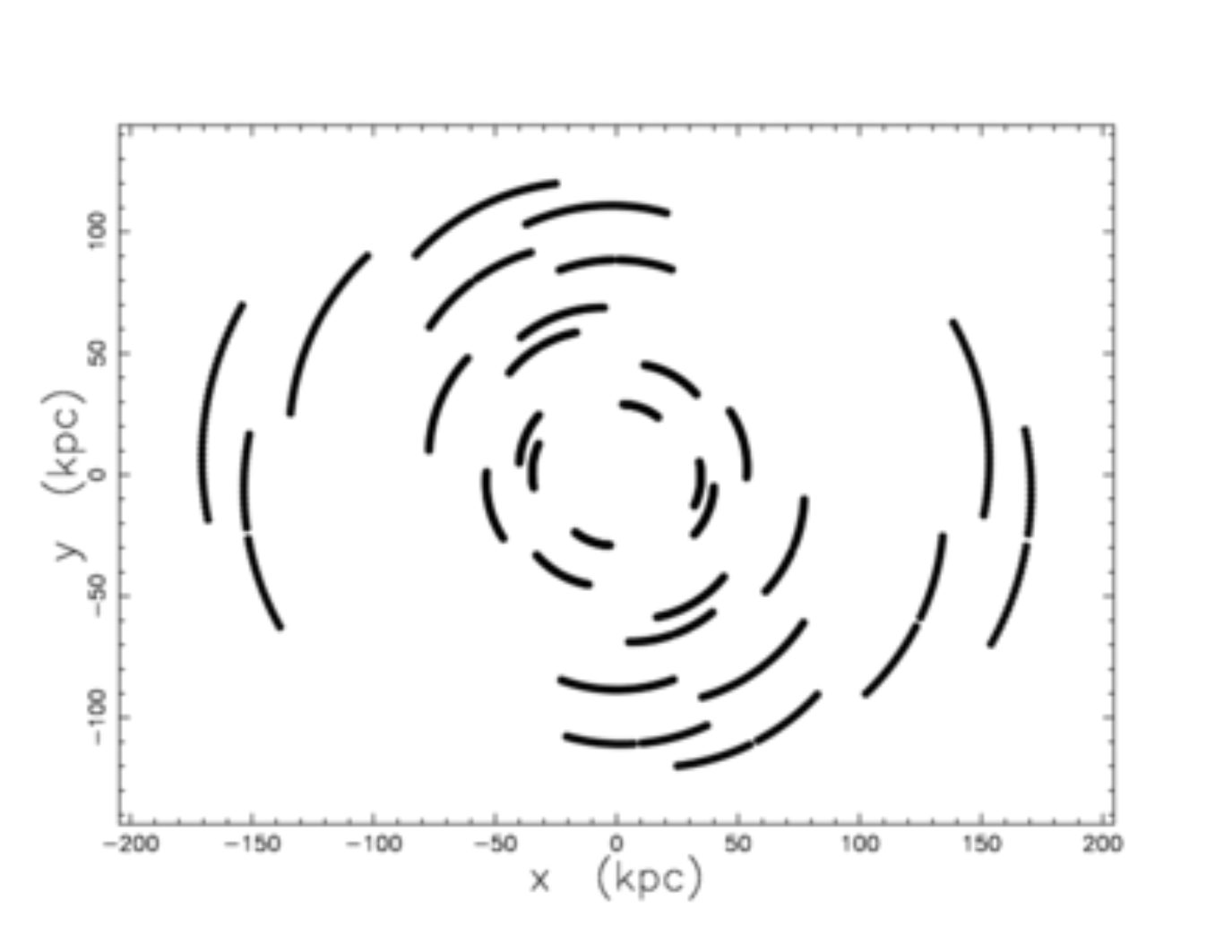}
\end{center}
\caption
{
Multiple sections of the SB as in Figure~\ref{allsb} 
with
angular extension  of the 
polar angle, $\theta$, 
of $\approx~31^{\circ}$ and progressive rotation  of the selected piece
of section. 
}
\label{allzoom}
\end{figure}

\begin{figure}
\begin{center}
\includegraphics[width=8cm,angle=-90]{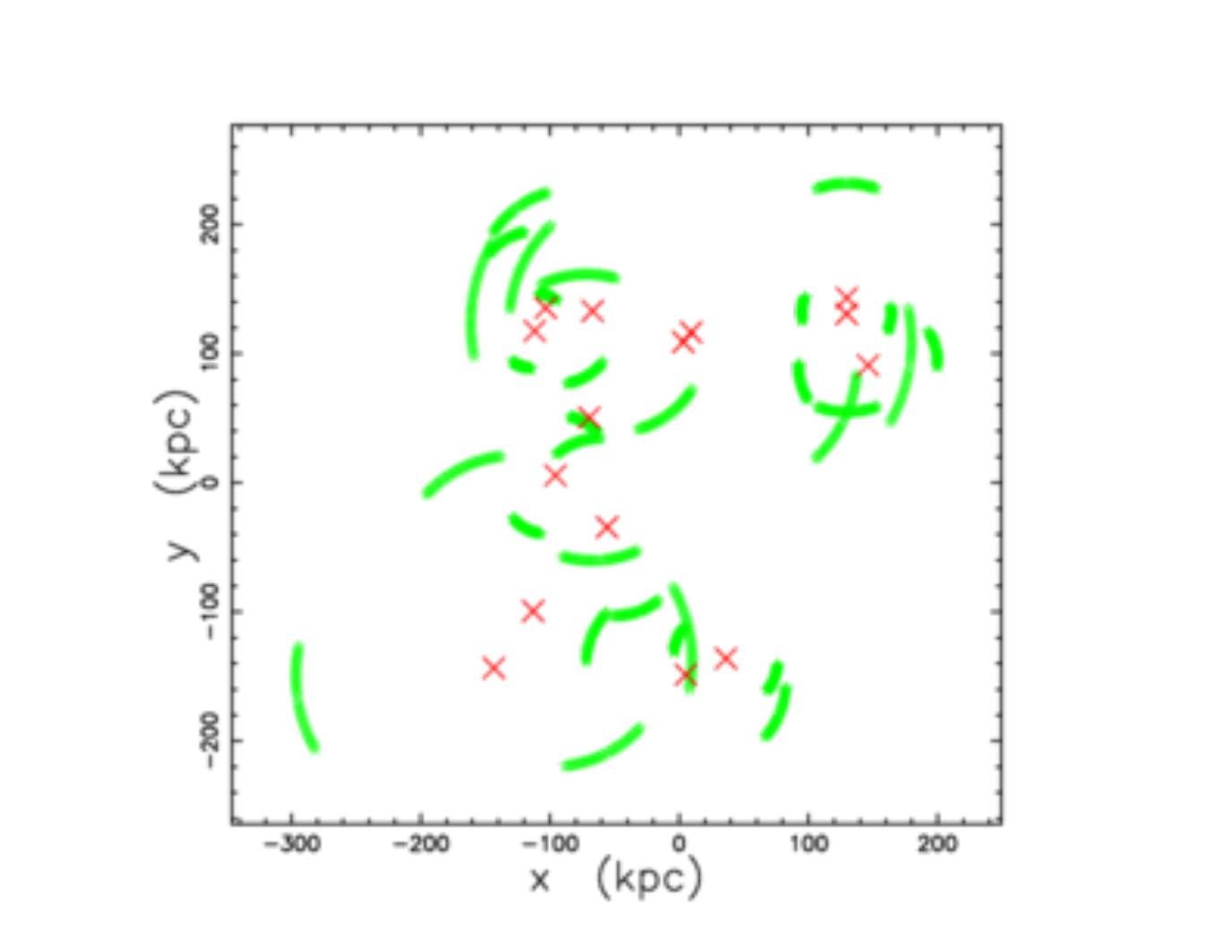}
\end{center}
\caption
{
Multiple sections of SB as in Figure~\ref{allzoom} 
with 
random shift of the origin of the selected 
SB (green empty stars).
The random shift denotes the galaxies (red crosses). 
}
\label{allastro}
\end{figure}

The presence of multiple  giants arcs 
in the CLASH cluster, see as an example Figure 11 in \cite{Xu2016},
can be simulated adopting the following steps
\begin{itemize}
\item 
A given number of SBs, as an example 15,  are generated
with variable lifetime, $t$, 
see Figure \ref{allsb}
\item 
For each SB we select a section around polar
angle equal to zero  
characterized  by a fixed angle of
$\approx~31^{\circ}$  and we randomly rotate it 
around the origin, see Figure \ref{allzoom} 
\item
The centers of the SBs  are randomly 
placed in  a  squared box
with side of 300~kpc, see Figure~\ref{allastro}
\end{itemize} 

Table \ref{center_center_theoretical} reports the  
theoretical statistical parameters  of the above simulation for the 
radial distance from the arc center to the cluster center in kpc.
A comparison should be done with the astronomical
parameters for the CLASH clusters  of Table~\ref{center_center}.

\begin{table}[ht!]
\caption
{
Statistical parameters of the
radial distance from the theoretical arc center 
to the cluster center in kpc  
}
\label{center_center_theoretical}
\begin{center}
{
\begin{tabular}{|c|c|c|}
\hline
 50 &  165  & 349    \\
\hline
\end{tabular}
}
\end{center}
\end{table}

\section{Theory of the image}

\label{imagetheory}

We now review the theory of the image 
for  the case of optically thin medium
both from an analytical 
and an analytical point of view.

\subsection{The elliptical shell}

A real ellipsoid
represents a first approximation of the 
asymmetric giants arcs
and has equation 
\begin{equation} 
\frac{z^2}{a^2} + \frac{x^2}{b^2} + \frac{y^2}{d^2}=1 
\quad ,
\label{ellipsoid}
\end{equation}
in which the polar axis    is the z-axis.

\begin{figure*}
\begin{center}
\includegraphics[width=8cm]{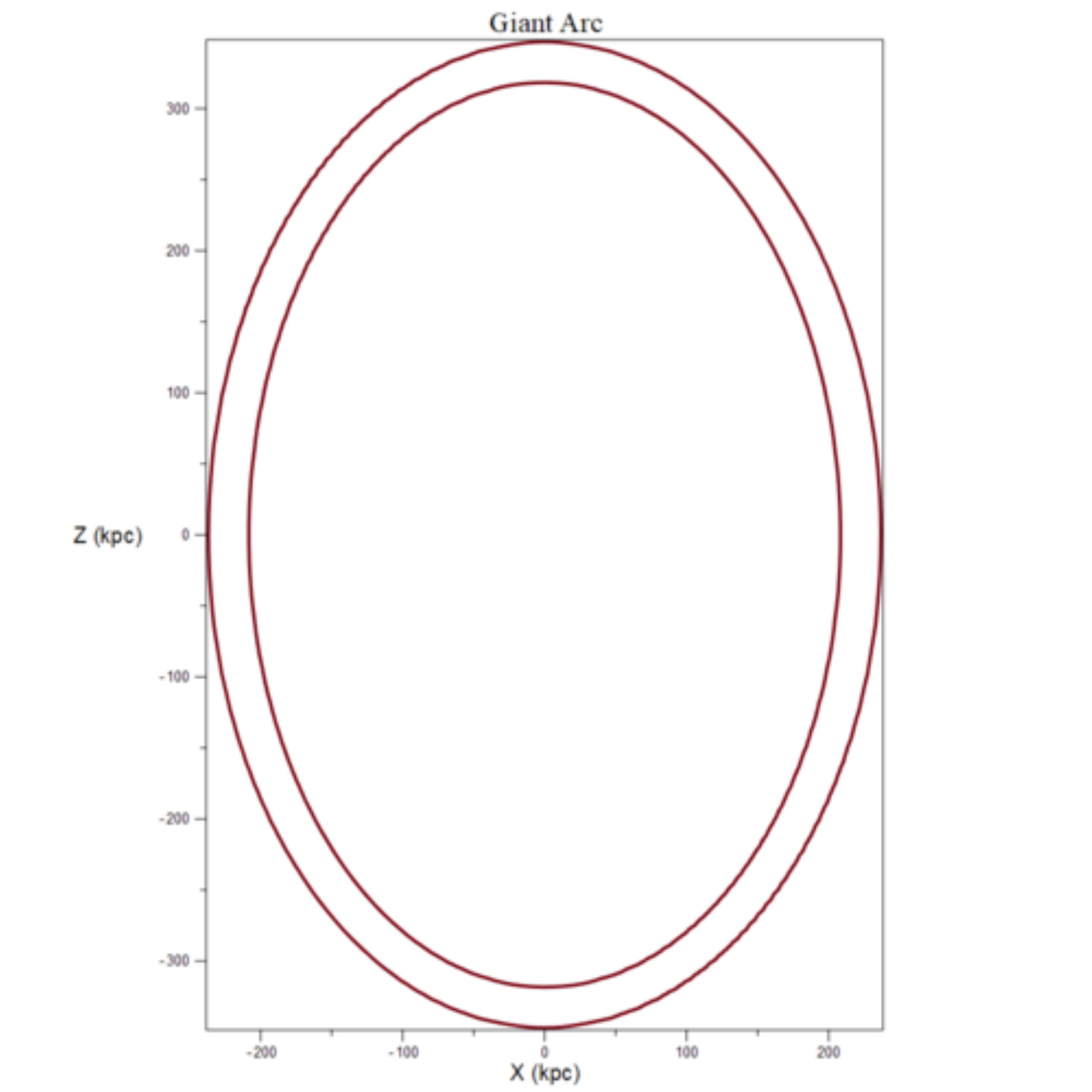}
\end {center}
\caption
{
Internal and external ellipses  
when $a=347\,kpc$,$b=237\,kpc$ and $c=\frac{a}{12}\,kpc$.
}
\label{int_ext_giantarc}
    \end{figure*}

\begin{figure*}
\begin{center}
\includegraphics[width=8cm,angle=-90]{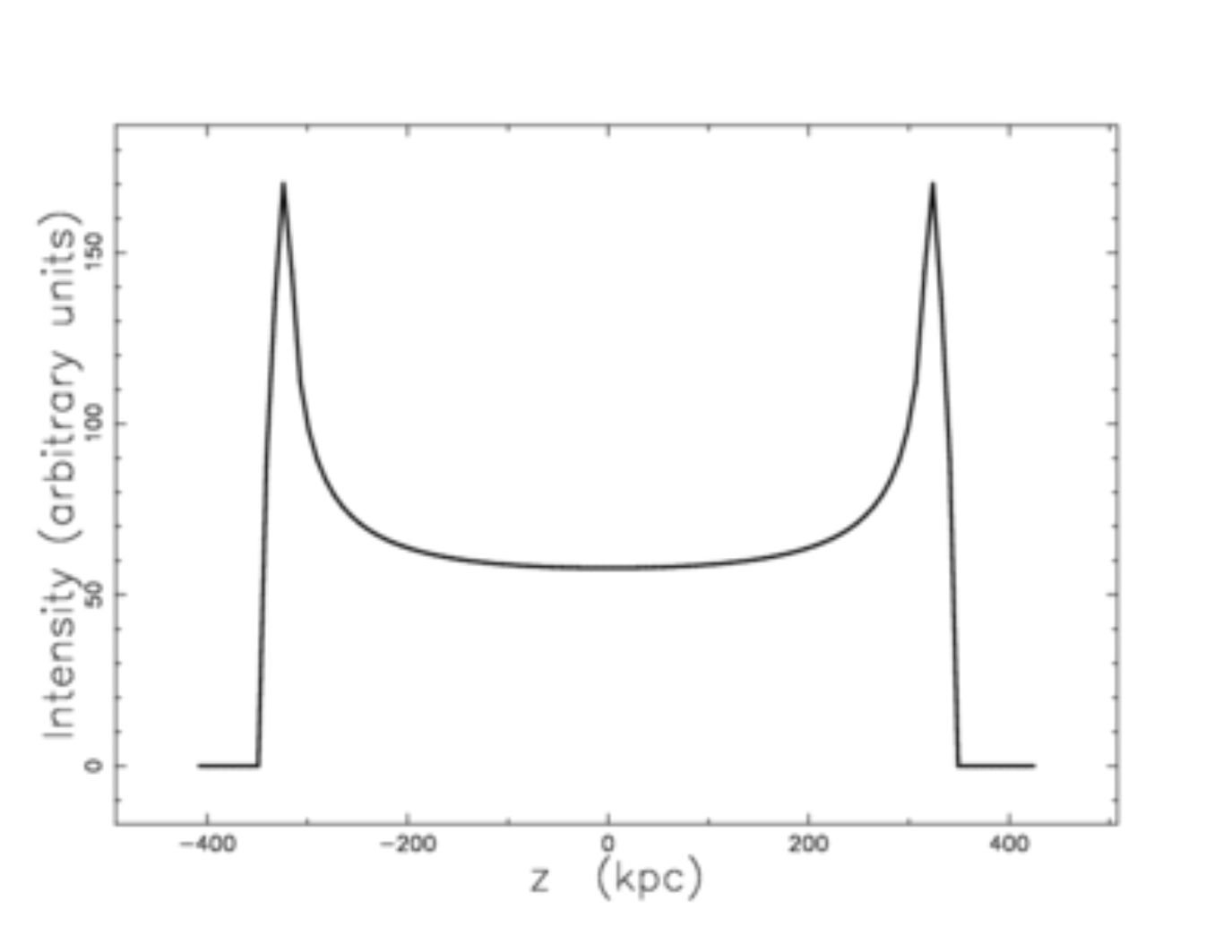}
\end {center}
\caption
{
The intensity profile along the z-axis when  
when $a=347\,kpc$,$b=237\,kpc$ $c=\frac{a}{12}\,kpc$
and $I_m$=1. 
}
\label{cut_ellipse_giant}
    \end{figure*}

We are interested in  the section of the ellipsoid $y=0$
which is defined by the following external ellipse
\begin{equation} 
\frac{z^2}{a^2} + \frac{x^2}{b^2} =1 
\quad .
\label{ellipse}
\end{equation}
We assume 
that the emission takes place in a  thin layer comprised between
the external  ellipse
and the  internal  ellipse defined by 
\begin{equation} 
\frac{z^2}{(a-c)^2} + \frac{x^2}{(b-c)^2} =1 
\quad ,
\label{ellipseint}
\end{equation}
see Figure \ref{int_ext_giantarc}.
We therefore
assume that the number density $C$ is constant and in particular
rises from 0 at  (0,a)  to a maximum value $C_m$, 
remains constant
up to (0,a-c)  and then falls again to 0. 
The length of sight, when
the observer is situated at the infinity of the $x$-axis, is the
locus parallel to the $x$-axis which  crosses  the position $z$ in
a Cartesian $x-z$ plane and terminates at the external 
ellipse.
The locus length
is
\begin{eqnarray}
l_I  =  2\,{\frac {\sqrt {{a}^{2}-{z}^{2}}b}{a}}
\\
when \quad   (a-c) \leq z < a  \nonumber  \\
l_{II} =  
2\,{\frac {\sqrt {{a}^{2}-{z}^{2}}b}{a}}-2\,{\frac {\sqrt {{a}^{2}-2\,
ac+{c}^{2}-{z}^{2}} \left( b-c \right) }{a-c}}
\\
when \quad   
0 \leq z < (a-c)    \quad .
\nonumber  
\label{length}
\end{eqnarray}
In the case of optically thin medium, 
the  intensity  is split in two cases  
\begin{eqnarray}
I_I(z;a,b)  = I_m \times 2\,{\frac {\sqrt {{a}^{2}-{z}^{2}}b}{a}}
\\
when \quad   (a-c) \leq z < a  \nonumber  \\
I_{II}(z;a,,c) =            \nonumber     \\
I_m \times 
\Big ( 
2\,{\frac {\sqrt {{a}^{2}-{z}^{2}}b}{a}}-2\,{\frac {\sqrt {{a}^{2}-2\,
ac+{c}^{2}-{z}^{2}} \left( b-c \right) }{a-c}}
\Big )
\\
when \quad   
0 \leq z < (a-c)    \quad ,
\nonumber  
\label{intensitycut}
\end{eqnarray}
where  $I_m$  is  a constant which  allows to compare the theoretical
intensity  with the observed one.
A typical profile in intensity along the z-axis  
is reported in Figure \ref{cut_ellipse_giant}.
The ratio, $\kappa$, between the theoretical intensity 
at the maximum,   $(z=a-c)$,
and at the minimum, ($z=0$), 
is given by 
\begin{equation}
\frac {I_I(z=a-c)} {I_{II}(z=0)} =\kappa= 
{\frac {\sqrt {2\,a-c}b}{\sqrt {c}a}}
\quad .
\label{ratioteorrim}
\end{equation}
As an example the values $a=6\,kpc$,$b=4\,kpc$, $c=\frac{a}{12}\,kpc$ 
gives $\kappa=3.19$.
The knowledge of the above ratio from the observations
allows to deduce $c$
once  $a$ and $b$ are given by the observed morphology
\begin{equation}
c =
2\,{\frac {a{b}^{2}}{{a}^{2}{r}^{2}+{b}^{2}}}
\quad .
\end{equation}
The above analytical model explains  the 
hole in luminosity visible 
in the astrophysical shells such as 
supernovae and SBs.
More details can be found in \cite{Zaninetti2018c}.

\subsection{The numerical shell}

The source of luminosity is assumed here to be
the flux of kinetic energy, $L_m$,
\begin{equation}
L_m = \frac{1}{2}\rho A  V^3
\quad,
\label{fluxkineticenergy}
\end{equation}
where $A$ is the considered area, $V$ is the velocity 
and $\rho$ is the density.
In our  case $A=r^2 \Delta \Omega$,
where $\Delta \Omega$ is the considered solid angle 
and $r(\theta)$ the temporary radius
along 
the chosen direction .
The   observed luminosity along a given direction 
can  be expressed as
\begin{equation}
L  = \epsilon  L_{m}
\label{luminosity}
\quad  ,
\end{equation}
where  $\epsilon$  is  a constant  of conversion
from  the mechanical luminosity   to  the
observed luminosity.

We review the algorithm that allows 
to build the image,
see \cite{Zaninetti2013c}:
\begin{itemize}
\item An empty 
memory grid  ${\mathcal {M}} (i,j,k)$ which  contains
$NDIM^3$ pixels is considered
\item We  first  generate an
internal 3D surface of revolution  by rotating the ideal image
of  $360^{\circ}$
around the polar direction and a second  external  surface of revolution 
at a
fixed distance $\Delta R$ from the first surface. As an example,
we fixed $\Delta R = R/12 $, where $R$ is the
momentary  radius of expansion.
The points on
the memory grid which lie between the internal and external
surfaces are memorized on
${\mathcal {M}} (i,j,k)$ by  a variable integer
number   according to formula
(\ref{fluxkineticenergy})  and   density $\rho$ proportional
to the swept    mass.
\item Each point of
${\mathcal {M}} (i,j,k)$  has spatial coordinates $x,y,z$ which  can be
represented by the following $1 \times 3$  matrix, $A$,
\begin{equation}
A=
 \left[ \begin {array}{c} x \\\noalign{\medskip}y\\\noalign{\medskip}{
\it z}\end {array} \right]
\quad  .
\end{equation}
The orientation  of the object is characterized by
 the
Euler angles $(\Phi, \Theta, \Psi)$
and  therefore  by a total
 $3 \times 3$  rotation matrix,
$E$.
The matrix point  is
represented by the following $1 \times 3$  matrix, $B$,
\begin{equation}
B = E \cdot A
\quad .
\end{equation}
\item
The intensity map is obtained by summing the points of the
rotated images
along a particular direction.
\end{itemize}

\begin{figure*}
\begin{center}
\includegraphics[width=10cm,angle=-90]{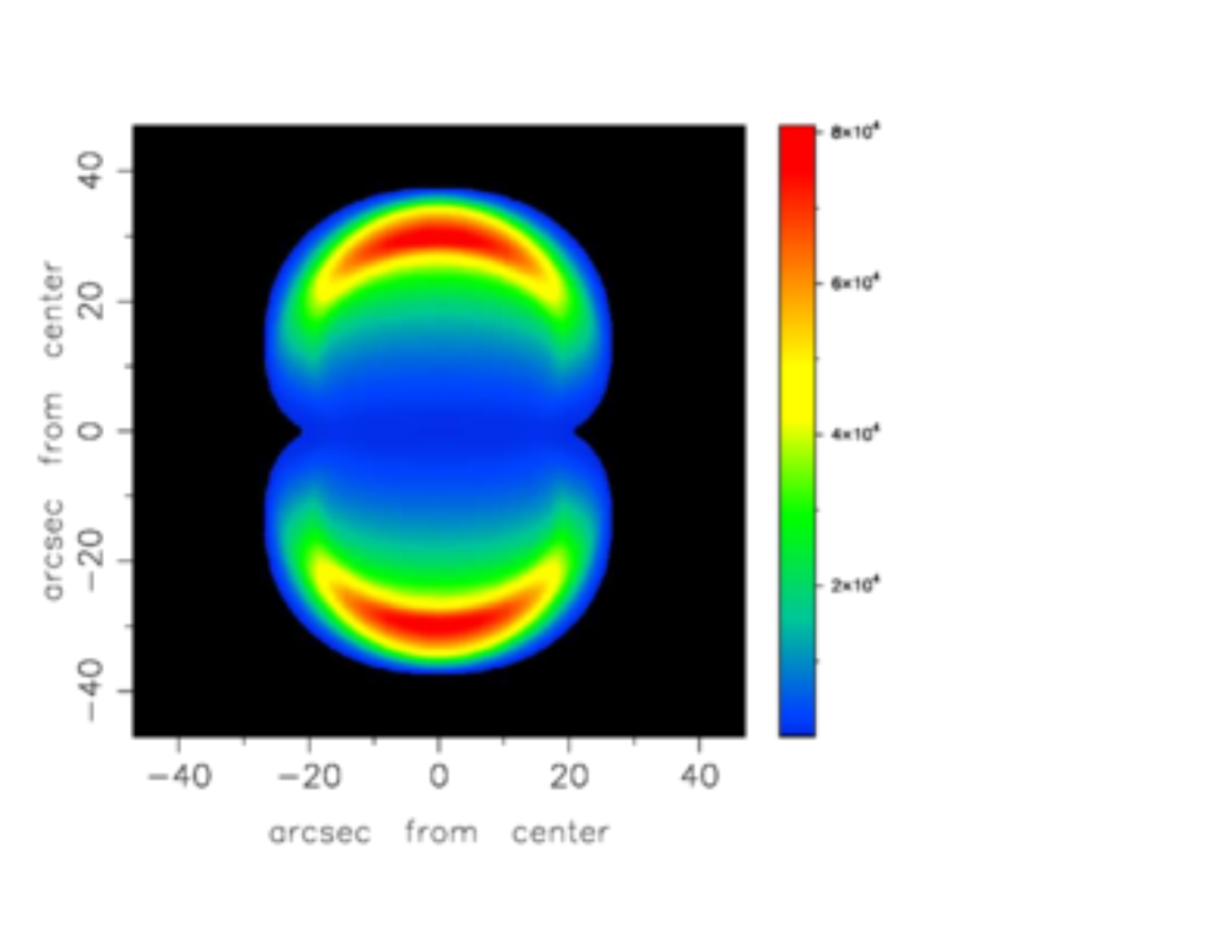}
\end {center}
\caption
{
Contour map  of  ${\it I}$ for A2267,
the $x$ and  $y$  axes  are in $arcsec$.
The three Euler angles
characterizing the   orientation
are $ \Phi $=0$^{\circ }$,
$ \Theta     $=90 $^{\circ }$
and   $ \Psi $=90  $^{\circ }$,
and NDIM=400.
}
\label{a2267_image}
    \end{figure*}

The image of A2267 built with the above algorithm 
is shown in Figure \ref{a2267_image}.

\begin{figure*}
\begin{center} 
\includegraphics[width=10cm,angle=-90]{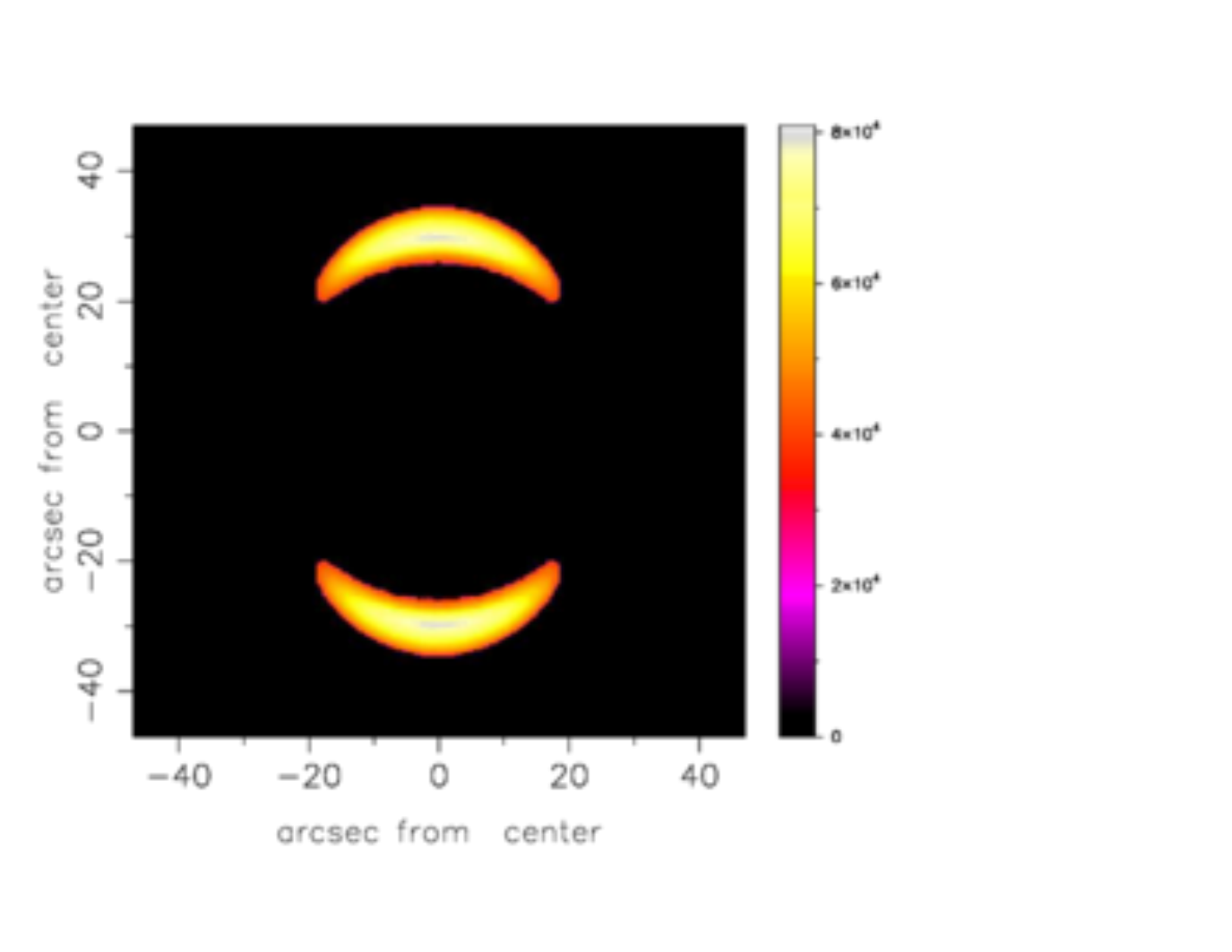}
\end {center}
\caption 
{   
The same  as  Figure \ref{a2267_image}  
parameters as in Figure \ref{a2267_image} and $\kappa=0.5$.
} 
\label{a2267_image_hole}
\end{figure*}

\begin{figure*}
\begin{center} 
\includegraphics[width=10cm,angle=-90]{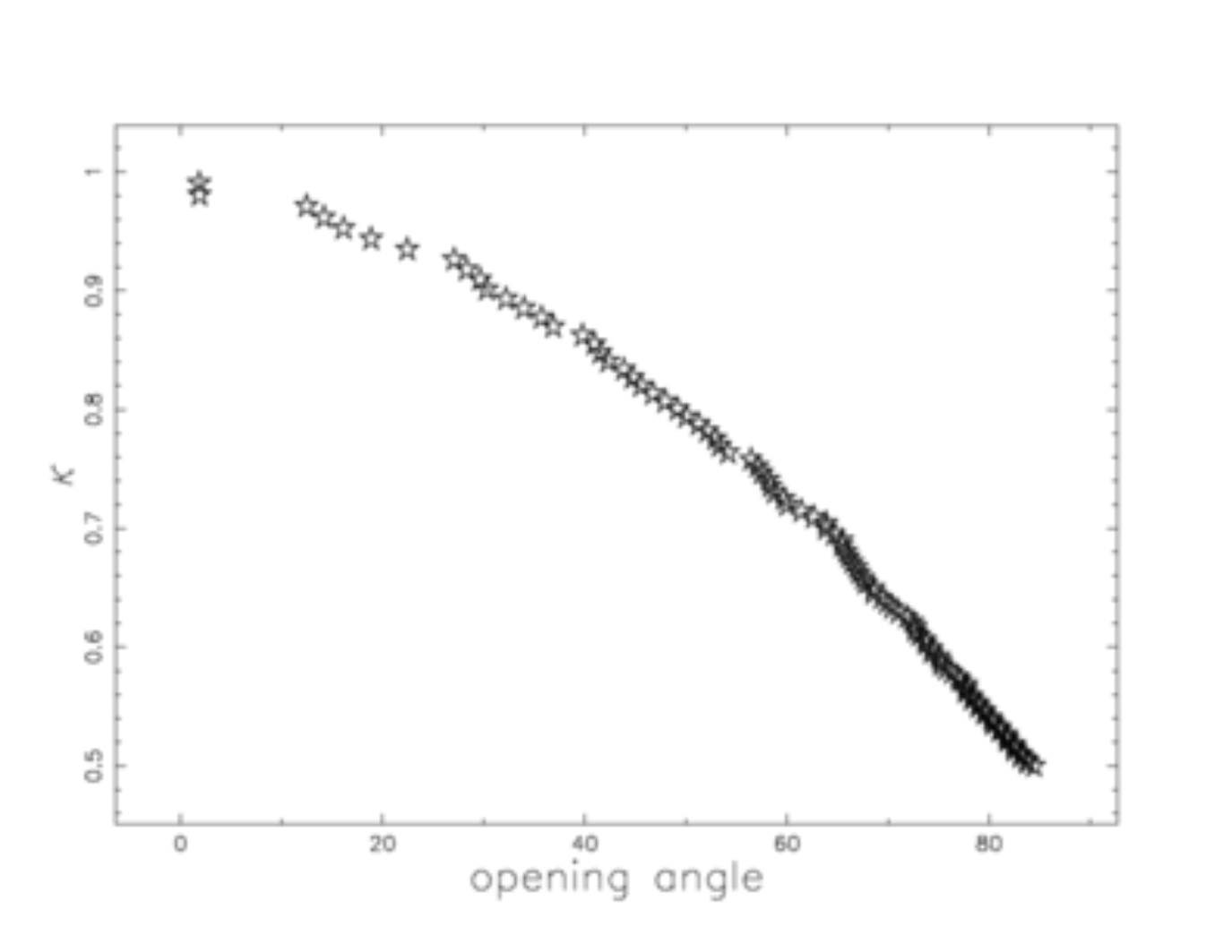}
\end {center}
\caption {   
The opening angle of the visible  simulated arc  as 
function of 
$\kappa$; 
parameters as in Figure \ref{a2267_image}.
} \label{a2267_image_opening}
    \end{figure*}

The
threshold intensity, 
$I_{tr}$,
is  
\begin{equation}
I_{max}\,\kappa= I_{max}
\quad ,
\end{equation}
where 
$I_{max}$, 
is the 
maximum  value  of intensity characterizing the ring
and  $\kappa$ is a parameter
which allows matching theory with observations
and was previously defined in equation~(\ref{ratiok}).
A typical
image with  a hole  is visible in  
Figure~\ref{a2267_image_hole}.
The opening angle of the visible arc 
can be parametrized as function 
of the ratio $\kappa$, see Figure 
\ref{a2267_image_opening}.
An opening of $\approx~31^{\circ} $  is reached  
at  $\kappa\approx 0.95$. 

\section{Conclusions}

{\bf The equation of motion}

The giants arcs  are connected with the visible part of 
the SBs   which advance in the intracluster 
medium  surrounding  the host galaxies.
The chosen profile  of density is hyperbolic,
see equation~(\ref{profhyperbolic}), 
and the momentum conservation along a given direction 
allows 
to derive the equation of motion
as function of  the polar angle,
see equation~(\ref{rtanalyticalhyper}).
\noindent
{\bf The  image}

According to the theory here presented 
the giants arcs are the visible part
of an advancing SB.
An analytical explanation for the limited 
angular  extent  of the giant arcs is represented 
by the theoretical luminosity as function of the 
polar angle, see equation~(\ref{ratiok}).
An increase in   the polar angle
produces a decrease of
the theoretical luminosity 
and the arc  becomes invisible.
Selecting a given numbers of SBs
with variable lifetime and randomly inserting them
in a cubic box of side $\approx~600\,kpc$ 
is possible to simulate
the giants arcs visible in the clusters of galaxies,
see Figure~\ref{allastro} 
and relative statistical  parameters in 
Table \ref{center_center_theoretical}.

\section*{Acknowledgments}

This research has made use of the VizieR catalogue access tool, CDS,
Strasbourg, France.

\providecommand{\newblock}{}

\providecommand{\newblock}{}

\end{document}